%% file: DBNR2021_arxiv.tex
\documentclass[12pt]{article}
\usepackage{amsmath}
\usepackage[utf8]{inputenc}
\usepackage[T1]{fontenc}
\usepackage[percent]{overpic}
\usepackage{amsfonts,amsthm,dsfont,enumerate}
\usepackage{graphicx}
\usepackage{enumerate, enumitem}
\usepackage{natbib} 
\usepackage[normalem]{ulem} 
\usepackage{url} 

\usepackage[ruled]{algorithm2e}
\SetKwInput{KwData}{Input}
\SetKwInput{KwResult}{Output}

\usepackage{colordvi}
\usepackage{graphicx}
 \usepackage{pgf, tikz,stmaryrd,subfig}
  \usetikzlibrary{fit,patterns,decorations.pathreplacing}
\usetikzlibrary{positioning,arrows.meta}

\newcommand{\blind}{1}

\newtheorem{theorem}{Theorem}[section]
\newtheorem{proposition}[theorem]{Proposition}

\newtheorem{lemma}[theorem]{Lemma}

\usepackage{xr-hyper}
\usepackage[bookmarks=false]{hyperref}

\newcommand{\R}{\mathbb{R}} 

\renewcommand{\P}{\mathbb{P}}

\newcommand{\Nm}{\mathbb{N}_m}

\newcommand{\FDPbar}{\overline{\mbox{FDP}}}
\newcommand{\marie}[1]{\textcolor{black}{#1}}

\newcommand{\pierre}[1]{\textcolor{black}{#1}}
\newcommand{\para}{\Gamma}

\newcommand{\UPI}{U^{\mbox{{\tiny PI}}}}
\newcommand{\Uboot}{U^{\mbox{{\tiny boot1}}}}
\newcommand{\Ubootbis}{U^{\mbox{{\tiny boot2}}}}

\newcommand{\Unnaive}{U^{\mbox{{\tiny naive}}}}
\newcommand{\Unrecentred}{U^{\mbox{{\tiny boot3}}}}

\newcommand{\Post}{\Pi}

\newcommand{\bell}{\boldsymbol{\ell}}

\newcommand{\E}{\mathbb{E}} 

\renewcommand{\l}{\ell}
\newcommand{\FDR}{\mbox{\rm FDR}}

\newcommand{\mtc}{\mathcal}

\newcommand{\wh}[1]{{\widehat{#1}}}

\newcommand{\ind}[1]{{\mathds{1}\{#1\}}}

\newcommand{\abs}[1]{\left| #1 \right|}

\newcommand{\cH}{{\mtc{H}}}

\usepackage{graphics}
\newcommand{\FDP}{\mbox{FDP}}

\SetKwProg{init}{Initialization}{}{}

\SetKwProg{Iteration}{iter}{}{}

\graphicspath{
  {../}
  {appli/}
  {Simu_papier/}
}

\begin{document}

\def\spacingset#1{\renewcommand{\baselinestretch}%
{#1}\small\normalsize} \spacingset{1}


\if1\blind
{  \title{Post hoc false discovery proportion inference under a Hidden Markov Model}
  
  \author{Marie Perrot-Dock\`es\thanks{ marie.perrot-dockees@u-paris.fr
     }\hspace{.2cm}
    \\
    Université de Paris, CNRS \\ MAP5 UMR 8145, Paris, France.\\
    and \\
    Gilles Blanchard\\
    Universit\'e Paris-Saclay, CNRS, Inria\\
    Laboratoire de math\'ematiques d’Orsay\\
    91405, Orsay, France.\\
    and \\
    Pierre Neuvial\\
    Institut de Math\'ematiques de Toulouse\\
    UMR 5219, Universit\'e de Toulouse, CNRS\\
    UPS, F-31062 Toulouse Cedex 9, France.\\
    and \\
    Etienne Roquain\\
    Laboratoire de Probabilit\'es, Statistique    et    Mod\'elisation,\\ Sorbonne Universit\'e, Universit\'e de Paris, CNRS, Paris, France.
    }
  \maketitle
} \fi

%

\bigskip
\begin{abstract}
  We address the multiple testing problem under the assumption that the
  true/false hypotheses are driven by a Hidden Markov Model (HMM), which is
  recognized as a fundamental setting to model multiple testing under dependence
  since the seminal work of \citet{sun2009large}. While previous work has
  concentrated on deriving specific procedures with a controlled False Discovery
  Rate (FDR) under this model, following a recent trend in selective inference,
  we consider the problem of establishing confidence bounds on the false
  discovery proportion (FDP), for a user-selected set of hypotheses that can
  depend on the observed data in an arbitrary way.  We develop a methodology to
  construct such confidence bounds first when the HMM model is known, then when
  its parameters are unknown and estimated, including the data distribution
  under the null and the alternative, using a nonparametric approach. In the
  latter case, we propose a bootstrap-based methodology to take into account the
  effect of parameter estimation error.  We show that taking advantage of the
  assumed HMM structure allows for a substantial improvement of confidence bound
  sharpness over existing agnostic (structure-free) methods, as witnessed both
  via numerical experiments and real data examples.
\end{abstract}


\spacingset{1.5} 


\section{Introduction}

\subsection{Context and motivation}\label{sec:motiv}

To analyze large, heterogeneous and complex data, the analyst often adopts a {\it post hoc} approach, by choosing methods, formulating questions, selecting models or variables,  on the basis of the data set. In particular, performing statistical inference after selection is a flourishing research field, often named as {\it selective inference}.
The main challenge is to avoid the selection bias, by properly calibrating the error probabilities or risks. 

\begin{figure}[!htp]
  \centering
  \includegraphics[width=0.99\textwidth]{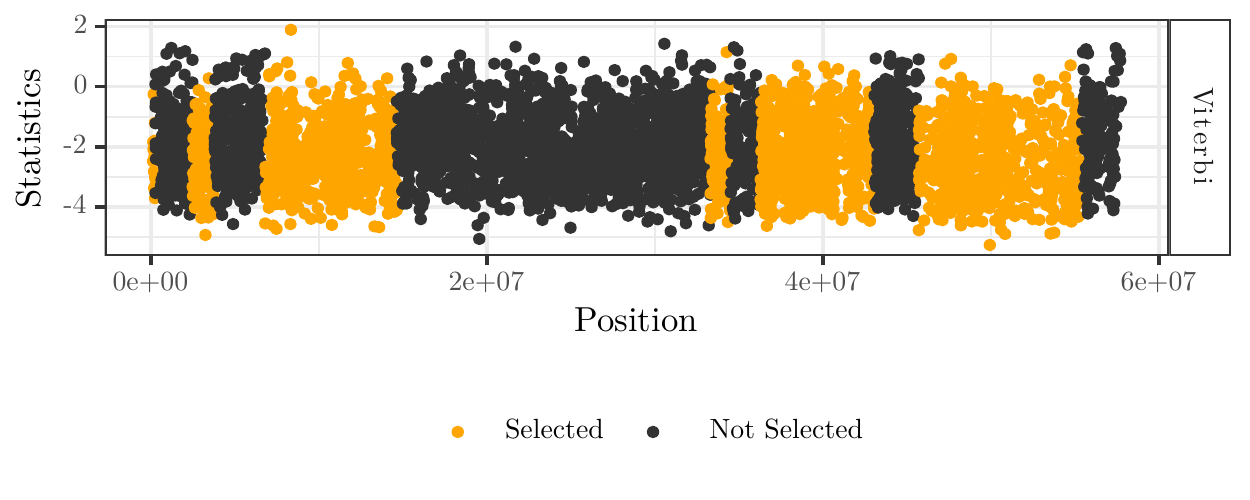}  
  \caption{Wilcoxon test statistics for $13,239$ loci on chromosome 7 in the Okamoto data set \citep{okamoto2015somatic}. Loci highlighted in orange correspond to a specific data-driven selection of regions.}
  \label{fig:okamoto-teaser}
\end{figure}

{Observations arising from applications are generally not independent. 
Hidden Markov Models (HMM) are a tool of choice to model stochastic processes with temporal or spatial dependence, and they have been widely successfully used in various areas including signal processing \citep{gales2008application} economics \citep{kim1999state}, or computational biology \citep{koski2001hidden}.
This paper is motivated by a specific use case in genomics: the differential analysis of DNA copy number alterations (CNA) in cancer cells. Cancer cells are characterized by structural changes in the number of gene copies along the genome, and modern biotechnologies such as microarrays and sequencing are commonly used to quantify such changes at high resolution~\citep{albertson2003chromosome}.  DNA copy number (CN)  profiles are generally modeled as piece-wise constant signals, and HMM have been extensively used for this purpose (see e.g. \citealp{fridlyand2004hidden,shah2009model,zhang2010dna}). 
Given the observation of CN profiles along the genome for individuals classified in two groups corresponding to different types of cancer, differential analysis aims at identifying regions of the genome for which the CN profiles differ ``significantly'' between the two cancer types. As a case in point, we consider a study of ovarian cancers where a differential analysis of $117$ patients with or without endometriosis is performed \citep{okamoto2015somatic}, see Section~\ref{sec:copy-numb-alter}. This study comprises CNA measurements for $236,385$ genomic locations (loci) for $63$ patients without endometriosis and $54$ patients with endometriosis. Figure~\ref{fig:okamoto-teaser} displays two-sided Wilcoxon test statistics of the null hypothesis of no signal difference between the two cancer types, for $4,799$ loci on chromosome~7. 
}
{We model these test statistics as an HMM with two states, corresponding to the null and alternative hypotheses for the test. Loci highlighted in orange correspond to a specific data-driven selection of regions. The methods developed in this paper make it possible to construct confidence bounds on the false positives in such data-driven selections. They are both more informative than approaches based on the control of local False Discovery Rates \citep{sun2009large}, which only provide posterior point estimates, and more powerful than agnostic post hoc bounds \citep{blanchard2020post} which do not take the dependency structure into account in the inference.}
{While the advantage of our approach will be put forward specifically with this data set (see Section~\ref{sec:copy-numb-alter}), comprehensive numerical experiments will show that this superiority holds in a wide spectrum of cases, see Sections~\ref{sec:num} and~\ref{sec:appli}; typically anytime assuming an HMM structure is reasonable.}

\subsection{Post hoc bounds}\label{sec:posthocbound}

The model considered here can be recast in the following more general setting: let us assume that the truth is carried by a {\em configuration vector} $\theta\in\{0,1\}^m$ (which may be random or not), for which $\theta_i=0$ if and only if the $i$-th variable is not active (in the above example $\theta_i=1$ then means chromosomal aberration at position $i$). Then, for an observation $X$, and any set $R\subset \Nm:=\{1,\dots,m\}$, we consider the following quantity of interest
\begin{align*}
\FDP(\theta,R)&=\frac{\sum_{i\in R} \ind{\theta_i=0}}{|R|\vee 1},
\end{align*}
which is called the false discovery proportion. Since its introduction by \citet{BH1995}, it has become a  classical way of accounting for the errors made by the selection $R=S(X)$ in a multiple testing context. 

Several ways of building confidence bounds for $\FDP(\theta,S(X))$ have been considered in the literature, {with varying interpretations and
degrees of generality for the possible selection policies $S(\cdot): x \mapsto S(x) \subset \Nm$ under
scrutiny.}
Confidence bounds valid simultaneously over all subsets $R\subset \Nm$ have been considered by \citet{genovese2006exceedance,goeman2011multiple,blanchard2020post}.  
Doing so, these bounds are also valid for any selection policy $S(\cdot)$.
A counterpart of this uniform guarantee is conservativeness, that is, the obtained bound can be far from the true value $\FDP(\theta,S(X))$ for a given particular selection policy $S(\cdot)$. Some improvements have been proposed in the literature by adding some local structure on the null hypotheses \citep{durand2020post} or on considering subset restricted to a ``path'' of procedures of interest \citep{KatRam2018}. 

Other approaches have been developed specifically in the linear model with various error rates concerning inferences after a model selection policy $S(\cdot)$ \citep{scheffe1959analysis,berk2013valid,bachoc2018post,tibetal2018,bachoc2019valid}. The LASSO policy is more specifically considered by \citet{lee2016exact}. Let us also mention that suitable confidence interval adjustments after selection have been proposed by \citet{BY2005,benjamini2014selective,WeinRamdas2019}. 

\subsection{Post hoc bounds in latent variables models}

{In this paper, we follow an approach very much in line with empirical Bayes methods for multiple testing, and in particular with the widely used two-group model and so-called ``local fdr'' method, see \citet{ETST2001,Efron2008, SS2018,JC2007,sun2009large,CS2009,CJ2010,heller2014,CSWW2019,RRV2019}, among others (more details about these studies can be found in Sections~\ref{sec:relptest} and~\ref{sec:relation}).}
To this end, the configuration
  vector $\theta$ is assumed to be a vector of random latent variables. We show that this modeling can greatly {simplify the
construction of} post hoc bounds, by using the distribution  of the latent variable conditionally on $X$, that we call the posterior distribution in analogy with Bayesian terminology. 

In the main part of the paper, following the seminal work of  \citet{sun2009large}, we will assume that $(\theta,X)$ follows an HMM structure, with model parameter $\Gamma$.
This model will be specified explicitly in Section~\ref{sec:model}.
{  For the general considerations in the remainder of the present section though, it is sufficient to assume
  that we have a model parametrized by $\para$ for the joint distribution 
   of $(\theta,X)$, denoted by $P_\Gamma$, typically specified through the marginal distribution of the configuration vector $\theta$, and the conditional distribution of the observation {$X$ conditionally on $\theta$}. The underlying distribution on the underlying probability space is denoted by $\P_\Gamma$ as usual.}

We consider the following aim: for any selection policy $S(\cdot):x\in \mathcal{X}\mapsto S(x)\subset\Nm$, build a functional $I_\alpha(X,S(\cdot))$ valued in the intervals of $[0,1]$, such that
\begin{align}\label{posthocbound3}
  \P_{\para} \left(\FDP(\theta,S(X)) \in I_\alpha(X,S(\cdot))\right)\geq 1-\alpha.
\end{align}
We call $I_\alpha(X,S(\cdot))$ a post hoc interval family with selection policy $S(\cdot)$.
Sometimes, the post hoc interval  $I_\alpha(X,S(\cdot))$ only requires knowledge of the selected region $S(X)$ on the observed data, 
in which case we will denote it by $I_\alpha(X,S(X))$ with an overload of notation.

{As a particular case, intervals of the form $I_\alpha(X,S(\cdot))=[0, \FDPbar_\alpha(X,S(\cdot))]$
can be considered, in which case \eqref{posthocbound3} reduces to a post hoc upper-bound $\FDPbar_\alpha(X,S(\cdot))$ on $\FDP(\theta,S(X))$. However,  \eqref{posthocbound3} also allows
for considering post-hoc lower-bounds,} which is also informative in practice, see Section~\ref{sec:appli}.

\subsection{Contribution: posterior post hoc bounds}
A standard approach in multiple testing to tackle~\eqref{posthocbound3} is to derive a bound which holds conditionally on any value $\theta$ of the latent true null hypothesis configuration, i.e.
  \begin{equation} \label{posthocboundusual}
    \P_{\para} \left(\FDP(\theta,S(X)) \in I_\alpha(X,S(\cdot)) \:|\:\theta\right)\geq 1-\alpha,
    \qquad \P_{\para}-\text{ a.s. }
  \end{equation}
  Note that this approach does not use at all the assumed model on the latent variables.
  To exploit this structural assumption, we consider instead
  solving \eqref{posthocbound3} via conditioning with respect to $X$ and solving
\begin{align}\label{posthocbound4}
  \P_{\para} \left(\FDP(\theta,S(X)) \in I_\alpha(X,S(\cdot)) \:|\:X\right)\geq 1-\alpha,
  \qquad \P_{\para}-\text{ a.s. , }
\end{align}
which can be achieved by considering the {\em posterior distribution}, that is, the distribution of $\theta$ conditionally on $X$ under $\P_\Gamma$.

If the parameter $\Gamma$ governing the distribution of the latent variables is known, one can build a posterior interval only depending on $S(X)$ that fulfills \eqref{posthocbound3}: denoting  $R=S(X)$, and provided $|R|>0$, 
\begin{align}
I_\alpha(X,R) &= [L_{\alpha\gamma}(X,R;\para),U_{\alpha(1-\gamma)}(X,R;\para)];\label{posteriorinterval}\\
U_{\beta}(X,R;\para) 
&= |R|^{-1} \min\left\{n\in \{0,\dots,m\}\::\: \P_\para\left(\sum_{i \in R} (1-\theta_i)\leq n\:\bigg|\: X\right)\geq 1-\beta\right\};\label{posteriorupperbound}\\
L_{\beta}(X,R;\para) 
&= |R|^{-1} \max\left\{n\in \{0,\dots,m\}\::\: \P_\para\left(\sum_{i \in R} (1-\theta_i)\geq n\:\bigg|\: X\right)\geq 1-\beta\right\},\label{posteriorlowerbound}
\end{align}
for some $\gamma\in(0,1)$ balancing the errors between the upper and lower bounds.
We will sometimes drop the $X$ in the notation for brevity.

This interval accounts for the particular HMM modeling via the posterior distribution, and thus
is expected to be considerably sharper than the other intervals described in Section~\ref{sec:posthocbound} and/or based on~\eqref{posthocboundusual}, that ignore this structure.
Unfortunately, the functionals $L_{\beta}(X,S(X);\para)$ and $U_{\beta}(X,S(X);\para)$ are not directly accessible because  the model parameter $\para$ is typically unknown.

We propose the following approximations of  $U_{\beta}(S(X);{\para})$ (similar for $L_{\beta}(S(X);{\para})$): 
\begin{itemize}
\item Plug-in:  $\UPI_{\beta}(X,S(X))=U_{\beta}(X,S(X);\wh{\para})$, where $\wh{\para}$ is an estimator of $\para$. We will consider an estimator $\wh{\para}$ based upon an iterative EM-type algorithm, see Section~\ref{sec:estim};
\item Bootstrap 1:  $\Uboot_{\beta}(X,S(\cdot))$,  correcting the fluctuations of the above plug-in bounds by using a bootstrap approach generating resampled data $X^*$ under $P_{\wh{\para}}$. This
bootstrap process requires the knowledge of the whole selection policy
$S(\cdot)$,  see Section~\ref{sec:boot1}. 
\item Bootstrap 2: $\Ubootbis_{\beta}(X,S(X))$, a heuristic approximation of
  $\Uboot_{\beta}(X,S(\cdot))$, which follows the same scheme, except that only
  $\wh{\para}$ is resampled; in particular the selection set $S(X)$ is kept
  fixed during the bootstrap process.  As a result, it does not require
  knowledge of the full selection policy $S(\cdot)$ but only of $R=S(X)$, the value of
  the selection policy for the observation $X$, see Section~\ref{sec:boot2}.
\item Bootstrap 3: $\Unrecentred_\beta(X,S(\cdot))$, a bootstrap bound based on generating resampled data $(\theta^*,X^*)$ under $P_{\wh{\para}}$ to approximate the distribution  of $\FDP(\theta,S(X))$, recentered with the  plug-in bound $U_{\beta}(X,S(X);\wh{\para})$, see Section~\ref{sec:boot3}.
  This process also requires the knowledge of the full selection policy.
\end{itemize}

These bounds have been implemented in the R package \texttt{SansSouci}, which is available from \url{https://pneuvial.github.io/sanssouci/}.

The coverage of these bounds is evaluated via extensive numerical experiments in Section~\ref{sec:num}, which typically reflect the impact of the estimation error of $\para$ in the different bounds.
Our conclusion is that while the existing approaches, ignoring the latent HMM structure, are over-pessimistic, the plug-in approach can be slightly over-optimistic. A good trade-off is provided by bootstrap-based strategies, which ensure a correct coverage while taking full advantage of the HMM structure.

\subsection{Relation to posterior point estimates of the FDR}
\label{sec:relptest}

In {previous literature, under a joint latent configuration/observation
  variable model $P_\para$ such as the one
considered here,
a commonly considered goal is to focus on the
the FDR of a given selection  policy $S(\cdot):x\in \mathcal{X}\mapsto S(x)\subset\Nm$}, that is,
\[
\FDR(S(\cdot),\para)= \E_{(\theta,X)\sim P_\para}( \FDP(\theta,S(X))).
\]
{Observe that,
since $\FDR(S(\cdot),\para)= \E_\para (\E_\para( \FDP(\theta,S(X))\:|\:X))
$, the quantity 
\begin{equation}\label{equFDRestimate}
\widehat{\FDR}(S(X),\para)=\E_\para( \FDP(\theta,S(X))\:|\:X)=|S(X)|^{-1}\sum_{i\in S(X)} \P_\para(\theta_i=0\:|\: X) 
\end{equation}
is an unbiased estimator for $\FDR(S(\cdot),\para)$, for any selection policy $S(\cdot)$.}

The marginal conditional probabilities $\P_\para(\theta_i=0\:|\: X)$, $1\leq i \leq m$, are generally referred to in the literature as the local FDR \citep{Efron2008} or the $\ell$-values \citep{CR2020}, but without the explicit purpose of being applied to any selection policy to our knowledge.
In fact, this  quantity can be used for several purposes:
\begin{itemize}
\item {for {\em any given} selection policy $S(\cdot)$, to compute an unbiased estimate of the
  FDR. Observe that this estimate only depends on the selected region $S(X)$, and not
  on the full selection policy $S(\cdot)$.}
\item {to {\em design} a selection policy $S(\cdot)$ such that $\FDR(S(\cdot),\Gamma)$ is equal to
  a specified level $\alpha$. The simplest way to achieve this is to construct $S(X)$ so that
  $\widehat{\FDR}(S(X),\para)$ is constant equal to $\alpha$. This is, for instance, the
    method proposed by \citet{SC2007,sun2009large,CS2009,CSWW2019}, where rejection regions taking the form
    of level sets of a relevant statistic $S_t(X)=\{i\in\Nm\::\: T_i(X)\geq t\}$
    are considered, and the threshold $t=t(X)$ is chosen also depending on $X$ as the
    smallest value so that $\widehat{\FDR}(S_{t(X)}(X),\para)=\alpha$.}
\end{itemize}
{In practice, the parameter $\Gamma$ is not known, so after introducing some suitable estimator $\wh{\para}$ of $\para$ (which can be built via an empirical Bayes approach), the
plug-in quantity $\widehat{\FDR}(S(X),\wh{\para})$ is used. In general, theoretical studies
then ensure that the above properties hold in a suitable asymptotical sense, see \citet{SC2007,sun2009large,CS2009,CSWW2019}.}

However, {this approach} only provides {an (asymptotically) unbiased} point estimate, and not a confidence interval for the FDP. The starting point of the present work is that we can
  use the same principle to derive confidence bounds for the FDP as in \eqref{posteriorinterval}-\eqref{posteriorupperbound}-\eqref{posteriorlowerbound},
  which are more informative for practice, see Figures~\ref{fig:ili2} and~\ref{fig:full7} in Section~\ref{sec:appli}. 
  Observe that, analogously to the two points above, the derived bounds can be used
  both to evaluate and design a policy.

\subsection{Relation to other works}
\label{sec:relation}

\paragraph{Multiple testing via a latent structure.}

Latent variables are widely used in statistics to design models with specific structures. In multiple testing, they have been used to model external ``confounding factors'' or ``systemic effects'' that induce dependencies between the tests, see \citet{leek2008general, friguet2009factor,fan2017estimation,fan2019farmtest} and references of \citet{Sunetal2012} for genetic or genomic applications.
In the present paper, the philosophy is rather different: latent variables are used  to model dependencies between the null hypothesis configurations (here, between the coordinates of the configuration vector $\theta$). While the unstructured case where the $\theta_i$ are i.i.d. can be traced back to the two-group model \citep{ETST2001,SC2007}, the structured HMM case has been shown to improve over the independent case by \citet{sun2009large}.  More recent examples of structures include two-sample sparsity \citep{CSWW2019} or  stochastic block-models for graph-structured nulls \citep{RRV2019}, for which substantial improvements are also shown with respect to the unstructured case.

\paragraph{Empirical Bayes FDR methods.}

Our method, as all methods cited above, has a Bayesian flavor, which is the case in general for all latent-based multiple testing based on the use of ``local fdr'', that is, the posterior probability that an item was generated under the null. While oracle versions (with a known model parameter) of such methods are known to be optimal is some way (see \citealp{SC2007} or more recently \citealp{heller2021optimal}), a challenging part is to evaluate how the methods can handle parameter estimation, which is generally shown in an asymptotic manner. As already underlined in Section~\ref{sec:relptest}, all these methods have been developed for the FDR metric, not for FDP confidence intervals. Since the FDP metric considered in our post hoc bounds is considerably more difficult to analyze than the FDR one, 
obtaining a consistency result for the coverage of our bounds is out of scope here. It is left as a challenge for future investigations, see Section~\ref{sec:consistency} for a discussion on this issue.

\paragraph{Non parametric inference in HMM.}

As said above, the feasibility  of parameter estimation plays a crucial role in empirical Bayes multiple testing. The quality of estimation heavily relies on the structural assumptions made for the latent variable distribution and for the considered model for the observations conditionally on the latent variables. Adding structure allows to increase performance. For an HMM with non parametric emission densities, the situation is more favorable than in the unstructured two-group model, see \citet{gassiat2016inference,alexandrovich2016nonparametric}, with provable consistency guarantees for the ``local fdr'' values \citep{de2017consistent,abraham2021multiple}.

\paragraph{Null estimation.}

Estimating the null density is both crucial and difficult in the two group model, as shown empirically by \citet{Efron2004, Efron2007b,  Efron2008,Efron2009b} and further studied by \citet{Sch2010, AS2015, Ste2017,SS2018}. In particular, the recent work of \citet{roquain2020false} theoretically shows that some sparsity is needed to obtain valid FDR inference when the null is estimated. By contrast, and as said above, the HMM assumption made in our setting makes this estimation much easier. In order to stick with the usual multiple testing line of research, we will make a distinction between the situation where the null distribution is assumed to be known (which is studied in the core of this paper as it is the most standard case), and the case where it is not (which is studied more specifically in Appendix~\ref{estimf0}). The numerical experiments of Section~\ref{sec:num} confirm the validity of our bounds even in the latter case.

\section{Hidden Markov modeling}\label{sec:model}

In this section, we define the HMM modeling of our problem and add some more material that will be useful to deal with the post hoc bounds.

\subsection{Model, notation and posterior distribution}\label{sec_model}

We use here a setting close to the model of \citet{sun2009large}.
Let us consider a hidden Markov process $(\theta,X) = ((\theta_i,X_i)_{1\leq i\leq m})$, where
\begin{itemize}
\item $\theta= (\theta_1,\dots,\theta_m)\in \{0,1\}^m$ is a unobserved latent variable sequence following a stationary Markov chain with matrix transition
$$
A =\left( \begin{matrix}
a_{0,0} & a_{0,1} \\
a_{1,0} & a_{1,1}
\end{matrix} \right),
$$
where $a_{q,\l}\in(0,1)$ for $q,\l\in\{0,1\}$, $a_{q,0} + a_{q,1} = 1$ for  $q \in\{0,1\}$ with $a_{0,0}\neq a_{1,0}$ to ensure that $A$ is full rank.
\item conditionally on $\theta$,  the observation $X=(X_i)_{1\leq i\leq m}\in \R^m$ has independent coordinates and for all $i\in\{1,\dots,m\}$,
$$
X_i\:|\: \theta_i \sim \left\{\begin{array}{cc}
f_0 & \mbox{ if $\theta_i=0$}\\
f_1 & \mbox{ if $\theta_i=1$}
\end{array}\right.,
$$
where $f_0$ and $f_1$ are two distinct densities on $\R$ (with respect to the Lebesgue measure).
\end{itemize}

Following \citet{gassiat2016inference}, this model is identifiable (which is a remarkable result since
$f_0,f_1$ can be arbitrary densities; it is a nonparametric model).
Moreover, since the sequence $\theta$ is assumed to be stationary, this implicitly means that $\theta_1$ is generated according to the marginal stationary distribution {$\pi$ on $\{0,1\}$} associated to $A$. Also, since we consider a testing paradigm where $f_0$ is the null distribution whereas $f_1$ is the alternative distribution, the roles of $f_0$ and $f_1$ are not exchangeable. We will consider both of the following cases:
\begin{itemize}
\item $f_0$ is known, in which case the parameter is $\para=(A,f_1)$. In that case, the multiple testing task aims at finding observations
  $X_i$ that depart from the distribution $f_0$;
\item $f_0$ is unknown, in which case the parameter is $\para=(A,f_0,f_1)$. In this case, since we do not know the distribution of the nulls, the task is detect `outliers', that are $X_i$'s with abnormal behavior. 
To distinguish $f_0$ from $f_1$, a classical assumption is $
\P(\theta_1=0) >\P(\theta_1=1)
$, that is, $f_0$ is the predominant class in the sample $X$. 
\end{itemize}

In our model, the distribution of $\theta$ conditionally on $X$ (the posterior distribution), is well known and is based on a quantity $\bell_{1, 1}(\para)$ and transition matrices $\Post_i(\para) $, $2\leq i \leq m$, whose explicit values are deferred to Appendix~\ref{sec:post}. 

\begin{proposition}[Proposition~3.3.2 of \citealp{cappe2006inference}]\label{prop:thetacondX}
In the model described in Section~\ref{sec_model}, the distribution of $\theta \:|\: X$ is a heterogeneous Markov Chain with transition matrices $\Post_i(\para) $, $2\leq i \leq m$ given by \eqref{equATX} and with initial distribution a Bernoulli distribution with parameter $\bell_{1, 1}(\para)$ given by \eqref{equlti}. 
\end{proposition}

\subsection{Parameter estimation}\label{sec:estim}

Since our posterior bounds depend on the unknown parameter $\Gamma$, building an estimator $\wh{\para}=(\wh{A},\wh{f}_1)$ of $\para=(A,f_1)$ is crucial to obtain a computable bound. In this section, we only deal with the case where $f_0$ is known, the other case being similar and postponed to Appendix~\ref{estimf0}.

For this, we make use of a pseudo Expectation-Maximization (EM) algorithm, combined with a  weighted kernel estimator of $f_1$, as proposed by \citet{robin2007semi} and studied by \citet{matias2014nonparametric}.  In our framework, this estimator can be written as follows: starting from an initial guess $\wh{\Gamma}^{(t-1)}$ of $\Gamma$, let
\begin{equation}\label{eq:estf1}
\wh{f}^{(t)}_1(x) = \sum_{i=1}^m \bell_{i, 1}(\wh{\para}^{(t-1)})\frac{K((x-X_i)/h)}{h} \bigg/ \sum_{i=1}^m \bell_{i, 1}(\wh{\para}^{(t-1)}),
\end{equation}
where  $K$ is a kernel function. We will typically use the Gaussian kernel $K(u)=(2\pi)^{-1/2}e^{-u^2/2}$, $u\in\R$. The rationale behind this is that weighting by $\bell_{i, 1}(\wh{\para}^{(t-1)})$ will put more weights to the observations $X_i$ for which $\theta_i$ is likely to be equal to $1$, that is, the observation under the alternative. Doing so, $\wh{f}^{(t)}_1(x)$ is close to an
{ideal} standard kernel 
estimator of the density of an i.i.d. sample that {would be based on the observations $X_i$ under the alternative only}.
This gives rise to Algorithm~\ref{algo:EM}. \\

\begin{algorithm}[H]
 \KwData{$X=(X_1,\dots,X_m)$}
 \KwResult{$\wh{\para}=(\wh{A},\wh{f}_1)$ estimator of $\para=(A, f_1)$}
 \begin{itemize}
 \item
  Initialization: first guess $\wh{\para}^{(0)}=(\wh{A}^{(0)},\wh{f}_1^{(0)})$ of $\para$
 \item
 Loop: at step $t\geq 1$, given $\wh{\para}^{(t-1)}$, do
 \begin{itemize}
 \item
 Compute 
\begin{enumerate}
\item $\alpha_i(q)=\alpha_i(q;\wh{\para}^{(t-1)})$ and $\beta_i(q)=\beta_i(q;\wh{\para}^{(t-1)})$, $1\leq i\leq m$, $q \in \{0,1\}$, using the forward-backward algorithm \eqref{eq:fwbc_al}, \eqref{eq:fwbc_bet};
\item $\bell_{i, q}(\wh{\para}^{(t-1)}) $, $1\leq i\leq m$, $q \in \{0,1\}$, 
by \eqref{equlti};
\item $\bell_{i, q,q'}(\wh{\para}^{(t-1)}) $, $1\leq i\leq m$, $q,q' \in \{0,1\}$, 
by \eqref{equltij}.
\end{enumerate}
 \item
Update
\begin{enumerate}
\item $\wh{a}^{(t )}_{q,q'} = \sum_{i = 1}^{m-1}  \bell_{i, q,q'}(\wh{\para}^{(t-1)}) /  \sum_{i = 1}^{m-1}  \bell_{i, q}(\wh{\para}^{(t-1)})$ , for $ 0 \leq q,q' \leq 1$;
\item 
$
\wh{A}^{(t )}  =\left( \begin{matrix}
\wh{a}^{(t )}_{0,0} & \wh{a}^{(t )}_{0,1} \\
\wh{a}^{(t )}_{1,0} & \wh{a}^{(t )}_{1,1}
\end{matrix} \right);
$
\item $\wh{f}_1^{(t )}$ by \eqref{eq:estf1}; 
\item $\wh{\para}^{(t )} = (\wh{A}^{(t )}, \wh{f}_1^{(t )})$.
\end{enumerate}
 \item
Stop the loop if convergence, that is, $\wh{\para}^{(t )}$ close enough to $\wh{\para}^{(t -1)}$.
 \end{itemize}
 \item Return $\wh{\para}=\wh{\para}^{(t)} $
 \end{itemize}
 \caption{\label{algo:EM} EM-type algorithm to derive $\wh{\para}=(\wh{A},\wh{f}_1)$ }
\end{algorithm}

\section{Posterior post hoc confidence bounds}
\label{sec:ourposthocbounds}

We first present the oracle bound, that is, the one using the true model parameter $\para$. Then, we introduce several options for estimating this oracle, firstly based on a rough ``plug-in'' via the estimator $\wh{\para}$ and then refinements based on bootstrap approaches. Also, we only present formulae for the upper bounds $U_{\beta}$ for brevity. The corresponding lower bounds $L_\beta$ are similar and quickly described in Appendix~\ref{sec:lowerbounds}.

\subsection{Oracle bound}\label{sec:oraclebound}

In this section, we fix any non-empty selected set $R=S(X)=\{j_1,\dots,j_s\}\subset\Nm$, for some indices $1\leq j_1<\dots<j_s\leq m$ and $s=|R|$.
Here, while the case where $R$ consists of contiguous indices is typical, we consider the more general situations where $R$ is arbitrary.
For notation brevity, we let $\theta^R_t=\theta_{j_t}$ for $t\in\{1,\dots,s\}$. 

According to \eqref{posteriorupperbound}, we let
\begin{equation}\label{equVnew}
U_{\beta}(X,R;\para)=s^{-1} \min\left\{n\in \{0,\dots,m\}\::\: \P_\para\left(\sum_{t=1}^s (1-\theta^R_{t})\leq n\:\bigg|\: X\right)\geq 1-\beta
 \right\}.
\end{equation}

Proposition~\ref{prop:thetacondX} ensures that, conditionally on $X$,  $(\theta^R_{t})_{1\leq t\leq s}$ follows an heterogeneous Markov chain with initial probability $\bell_{j_1, 1}(\para)$
and transition matrices 
\begin{align}
\Post^R_t (\para)&= \prod_{i=j_{t-1}+1}^{j_t} \Post_i(\para) ,\:\:\:\:\: \Post^R_t(\para) = \left(
\begin{matrix}
\Post^R_{t,0,0}(\para)  & \Post^R_{t,0,1}(\para) \\
\Post^R_{t,1,0}(\para) & \Post^R_{t,1,1}(\para) \\
\end{matrix}
\right), \:\: t\in\{2,\dots,s\}.
\end{align}
As such, $U_{\beta}(X,R;\para)$ is not explicit. 
In the sequel, we provide an algorithm to compute $U_{\beta}(X,R;\para)$.
 For this, let for $1\leq k \leq s$, $0\leq \l\leq s$,
 \begin{align}
 B_{k,\l,0}&= \P_\para\left( \sum_{t=1}^k (1-\theta^R_{t})\leq \l , \:\theta^R_k=0\:\bigg|\: X\right)\label{equA0};\\
 B_{k,\l,1}&= \P_\para\left( \sum_{t=1}^k (1-\theta^R_t)\leq \l, \:\theta^R_k=1\:\bigg|\: X\right)\label{equA1}.
 \end{align}
 In words, $B_{k,\l,0}$ is the posterior probability that there are at most $\l$ zero-occurrences in $\theta^R_{1:k}$, with a zero in the last position. Similarly, $B_{k,\l,1}$  is the posterior probability that there are at most $\l$ zero-occurrences in $\theta^R_{1:k}$, with a one in the last position. Since $B_{s,\l,0}+B_{s,\l,1}$ is the  posterior probability that at most $\l$ zero-occurrences occurs in the whole sequence $\theta^R=\theta^R_{1:s}$, the following holds. 
  
 \begin{proposition}\label{propnewbound}
The quantity $U_{\beta}(X,R;\para)$ defined by \eqref{equVnew} can be computed as
 \begin{equation}\label{equVnew2}
U_{\beta}(X,R;\para)=s^{-1} \min\left\{n\in \{0,\dots,m\}\::\: 
   B_{s,n,0}+ B_{s,n,1}\geq 1-\beta
 \right\},
\end{equation}
where $B_{k,\l,0}$ and $ B_{k,\l,1}$ are defined by \eqref{equA0} and \eqref{equA1}, respectively.
\end{proposition}

We present below an explicit algorithm to compute the quantities $B_{k,\l,0}$ and $ B_{k,\l,1}$, see Algorithm~\ref{algo:Bkl}.
In words, \eqref{recursA1} comes from the fact that having at most $\l$ zero-occurrences in $\theta^R_{1:k}$ with a zero in the last position means that we have at most $\l-1$ zero-occurrences in $\theta^R_{1:(k-1)}$ with either a zero or a one in position $k-1$. As for \eqref{recursA2}, the fact that having at most $\l$ zero-occurrences in $\theta^R_{1:k}$ with a one in the last position means that we have at most $\l$ zero-occurrences in $\theta^R_{1:(k-1)}$ with either a zero or a one in position $k-1$. \\
\begin{algorithm}[H]
 \KwData{$\Post^R_k(\para)$, $1 \leq k \leq s$; $\bell_{j_k, q}(\para)$, $1 \leq k \leq s$, $q \in \{0,1\}$ }
 \KwResult{$B_{k,\l,0}$ and $B_{k,\l,1}$, $1\leq k \leq s$, $0\leq \l\leq s$}
 \init{
$ B_{k,0,0}  =0$, $1\leq k \leq s$, $B_{1,\l,0} =\bell_{j_1, 0}(\para),$ $ 1  \leq \l \leq s$, $ B_{1,\l,1} =\bell_{j_1, 1}(\para) ,$ $0 \leq \l \leq s$.
 }{}{}
  \For{$2  \leq k \leq s$}{
  $B_{k,0,1} = B_{k -1,0,1} \Post^R_{k,1,1}(\para)$\\
  \For{$1  \leq \l \leq s$}{
  \vspace{-5mm}
 \begin{align}
 B_{k,\l,0}&= B_{k-1,\l-1,0}\: \Post^R_{k,0,0}(\para)+B_{k-1,\l-1,1}\:\Post^R_{k,1,0}(\para);\label{recursA1} \\
 B_{k,\l,1}&= 
 B_{k-1,\l,0}\:\Post^R_{k,0,1}(\para)+B_{k-1,\l,1}\:\Post^R_{k,1,1}(\para).\label{recursA2}
 \end{align}
   }
   }
 \caption{Computation of $B_{k,\l,0}$ and $B_{k,\l,1}$. \label{algo:Bkl}}
\end{algorithm}

\subsection{Plug-in bound}

The first non-oracle bound that is proposed is simply obtained by plugging the estimator derived in the Section~\ref{sec:estim} into the oracle bound, that is,
\begin{equation}\label{Qplugin}
\UPI_\beta(X,S(X))=U_{\beta}(X,S(X);\wh{\para}), 
\end{equation}
where $\wh{\para}$ comes from Algorithm~\ref{algo:EM} and $U_{\beta}(X,R;\para)$ is given by \eqref{equVnew} and \eqref{equVnew2}. {Since the oracle bound is based on the conditional
  distribution of the latent variable given the observation, the above bound can be interpreted as an ``empirical Bayes
credible set'' for the FDP.}

Unfortunately, this plug-in bound can be anti-conservative, meaning that it can violate~\eqref{posthocbound3}, as we will see in the simulations of Section~\ref{sec:num}.  
{An intuitive explanation} is that 
$U_{\beta}(X,S(X);\wh{\para})$ is expected to fluctuate on both sides of $U_{\beta}(X,S(X);\para)$.
and thus the event $\FDP(\theta,S(X))\leq U_{\beta}(X,S(X);\wh{\para})$ may not be true when $U_{\beta}(X,S(X);\wh{\para})$ is smaller than $U_{\beta}(X,S(X);{\para})$.

In other words, the variability w.r.t. {the posterior distribution of $\theta$ conditional on $X$} is taken into account by $U_{\beta}$, but it is still needed to evaluate the variability in $X$ of the bounds, which also relies both on the variability of the estimate $\wh{\para}$ of $\para$ and the policy $S(\cdot)$. Our first bootstrap bound takes both variations into account, while the second one focuses solely on the  variation of $\wh{\para}$.

\subsection{First bootstrap bound}\label{sec:boot1}

Let us consider the deterministic quantity
\begin{equation}\label{equ-qgamma}
q_{1,\gamma}(\beta,S(\cdot);\para)= \min\left\{x\in \R \::\: \P_\para\left(U_{\beta}(X,S(X);{\para})- U_{\beta}(X,S(X);\wh{\para}) \leq x\right)\geq 1-\gamma\right\}
\end{equation}
which corresponds to the $(1-\gamma)$-quantile of  the distribution of $U_{\beta}(X,S(X);\para)- U_{\beta}(X,S(X);\wh{\para})$ when $X$ is generated according to the {true} model parameter $\Gamma$. Then, we have by definition, for all $\delta\in (0,1)$,
\begin{align*}
\P_\para( \FDP(\theta,S(X))\leq U_{\beta(1-\delta)}(X,S(X);\para)\:|\: X)&\geq 1-\beta(1-\delta)\\
\P_\para(U_{\beta(1-\delta)}(X,S(X);{\para})-U_{\beta(1-\delta)}(X,S(X);\wh{\para})\leq q_{1,\beta\delta}(\beta(1-\delta),S(\cdot);\para))&\geq 1-\beta\delta.
\end{align*}
Note that the first bound concerns the distribution of $\theta$ conditionally on $X$ while the second one concerns only the marginal distribution of $X$.
Therefore, this immediately implies a bound with respect to the joint distribution of $(\theta,X)$.

\begin{proposition}
  For any selection policy $S(\cdot):x\in \mathcal{X}\mapsto S(x)\subset\Nm$ and $\delta\in (0,1)$, for any model parameter $\Gamma$,  we have 
 \begin{equation}\label{equ:covboot1}
 \P_\para(\FDP(\theta,S(X))\leq U_{\beta(1-\delta)}(X,S(X);\wh{\para})+ q_{1,\beta\delta}(\beta(1-\delta),S(\cdot);\para))\geq 1-\beta.
 \end{equation}
\end{proposition}

Hence, the RHS of the event in \eqref{equ:covboot1} is a post hoc bound in the sense of \eqref{posthocbound3}. However, again, it depends on the unknown parameter $\para$, although it is only via the quantity $q_{1,\gamma}(\beta,S(\cdot);\para)$, that we would like to think of as a ``second order'' term.
 
Now, plugging the estimate $\wh{\para}$ of $\para$ in the latter leads to the bootstrap-type approximation
$
q_{\gamma}(\beta,S(\cdot);\para)\approx q_{\gamma}(\beta,S(\cdot);\wh{\para})$ 
which corresponds to consider the $(1-\gamma)$-quantile of  the distribution of $U_{\beta}(X^*,S(X^*);\wh{\para})- U_{\beta}(X^*,S(X^*);\wh{\para}^*)$ under the model parameter $\wh{\para}$, {where $X^*$ is an independent sample generated under $P_{\wh{\para}}$ and
  $\wh{\para}^* = \wh{\para}(X^*)$}, see \eqref{equ-qgamma}. As usual, the bootstrap quantity $q_{\gamma}(\beta,S(\cdot);\wh{\para})$ is in turn approximated by a quantity $\tilde{q}^{(B)}_{1,\gamma}(\beta,S(\cdot);\wh{\para})$ obtained via a Monte-Carlo scheme that generates $B$ times the sequence $(X^*_i)_{1\leq i\leq m}$  according to the model with parameter $\wh{\para}$. 

We are now in position to introduce our first bootstrap bound:
\begin{equation}\label{Qboot1}
\Uboot_{\beta,\delta}(X,S(\cdot))=
U_{\beta(1-\delta)}(X,S(X);\wh{\para})+ \tilde{q}^{(B)}_{1,\beta\delta}(\beta(1-\delta),S(\cdot);\wh{\para})
,
\end{equation}
Algorithm~\ref{algo:boot_bounds} describes how this bound may be computed. 

Let us make some comments on this bound: 
\begin{itemize}
\item {This bootstrap bound has a semi-parametric flavor: while the $\theta$ sequence is generated via a parametric Markov chain using the estimated parameters, the resampled data  $X_i^*$ are obtained
    by weighted  smooth bootstrap \citep{EfronBootstrap1979}, since they
are drawn from     
    a weighted kernel density estimator (see Appendix~\ref{sec:app_impl_boot} for more details).}
\item The term $\delta$ balances the way the errors are distributed and should be chosen to derive an appropriate trade-off: a small $\delta$ will sharpen the bound $U_{\beta(1-\delta)}(X,S(X);\wh{\para})$ but makes $\tilde{q}^{(B)}_{1,\beta\delta}(\beta(1-\delta),S(\cdot);\wh{\para})$ larger.  In the numerical experiments, we have {observed that the impact of $\delta$ is moderate.}
\item In this bound, we should compute $S(X^*)$, see Algorithm~\ref{algo:EMf0unknown}. This means that the user should provide their whole selection policy $S:x\in \mathcal{X}\mapsto S(x)\subset\Nm$, and not only the selection $S(X)$ on the observed data set $X$. This can be seen as a constraint in  some cases, and the next paragraph provides a solution to circumvent this limitation.
\item Whenever $\tilde{q}^{(B)}_{1,\beta\delta}(\beta(1-\delta),S(\cdot);\wh{\para})$ is nonnegative (which happens with high probability for suitable values of $\beta$ and $\delta$), we have  $\Uboot_{\beta,\delta}(X,S(\cdot))\geq 
U_{\beta(1-\delta)}(S(X);\wh{\para})$ so the obtained bound is in general more conservative than the plug-in bound.
\end{itemize}
\begin{algorithm}[H]
 \KwData{{Common input for all the bounds : Data  $X$; Estimator $\wh{\para}$ via Algorithm~\ref{algo:EM} if $f_0$ is known (or by Algorithm~\ref{algo:EMf0unknown} if $f_0$ is unknown); $\beta\in (0,1)$
 \begin{itemize}
 \item {Input $\Uboot$: } Selection policy $S(\cdot):x\in \mathcal{X}\mapsto S(x)\subset\Nm$; , $\delta\in (0,1)$
 \item {Input $\Ubootbis$: } $\delta\in (0,1)$, $S(X)$ 
 \item  {Input $\Unrecentred$: } Selection policy $S(\cdot):x\in \mathcal{X}\mapsto S(x)\subset\Nm$
 \end{itemize}
 }
 }
 \KwResult{{
     $\Uboot_{\beta,\delta}( S(\cdot), \wh{\para})$ {or} $\Ubootbis_{\beta,\delta}( S(X), \wh{\para}) $ or $\Unrecentred_\beta( S(\cdot), \wh{\para})$} 
 }
\begin{enumerate}
\item Generate independently $B$ bootstrap samples as follows: for $1\leq b \leq B$
\begin{enumerate}
\item Draw $\theta^{*(b)}$ and $X^{*(b)}$ from $\wh{\para}$
%
  \item Compute $\wh{\para}^{*(b)}$ using Algorithm~\ref{algo:EM} with the sequence $X^{*(b)}$ (or using Algorithm~\ref{algo:EMf0unknown} if $f_0$ is unknown);
 \item Compute  {$D_i^{(b)}$, for the appropriate $i\in (1,2,3)$ and $\lambda\in (\beta(1-\delta),\beta(1-\delta),\beta)$};
  \begin{itemize}
 \item {$\Uboot$: }
 $D_1^{(b)} = 
 U_{\lambda}(X^{*(b)},S(X^{*(b)});\wh{\para})- U_{\lambda}(X^{*(b)},S(X^{*(b)});\wh{\para}^{*(b)})$;
 \item {$\Ubootbis$: }
 $D_2^{(b)} = 
 U_{\lambda}(X,S(X);\wh{\para})- U_{\lambda}(X,S(X);\wh{\para}^{*(b)})$;
 \item {$\Unrecentred$: }
  $D_3^{(b)} = 
 \FDP(\theta^{*(b)},S(X^{*(b)}))- U_{\lambda}(X^{*(b)},S(X^{*(b)});\wh{\para}^{*(b)})$,
\end{itemize}  
\end{enumerate}
where $U_{\lambda}(X,S(X);\para)$ is given by \eqref{equVnew} and \eqref{equVnew2}
\item Compute $\tilde{q}^{(B)}_{i,\gamma}(\lambda)$ as the empirical $\gamma$-quantile of ${D_i} = (D_i^{(1)},\dots, D_i^{(B)})$ {for the appropriate $\gamma\in (\beta\delta,\beta\delta,\beta)$}; 
\item 
{Return the corresponding bound}
\begin{itemize}
\item $\Uboot_{\beta,\delta}(X, S(\cdot), \wh{\para}) =
U_{\beta(1-\delta)}(X, S(X),\wh{\para})+ \tilde{q}^{(B)}_{1, \beta\delta}(\beta(1-\delta))$
\item $\Ubootbis_{\beta,\delta}(X, S(X), \wh{\para}) = 
U_{\beta(1-\delta)}(X, S(X),\wh{\para})+ \tilde{q}^{( B)}_{2, \beta\delta}(\beta(1-\delta))$
\item   $\Unrecentred_\beta(X, S(X), \wh{\para}) =
U_{\beta}(X, S(X),\wh{\para})+ \tilde{q}^{( B)}_{3,\beta}(\beta)$
\end{itemize}
\end{enumerate}
\caption{{Core algorithm for computing the bootstrap bounds} \label{algo:boot_bounds}\label{algo:qbeta}} 
\end{algorithm}

\subsection{Second bootstrap bound}\label{sec:boot2}

As mentioned above, computing $\Uboot_\beta(\cdot)$ requires that the user provides $S(X^{*(b)})$ for every single bootstrap sample $X^{*(b)}$, $b=1,\dots,B$.  This could be inconvenient if the user does not want to provide the whole selection policy, but only $S(X)$, the desired selection on the observed data set and not on other virtual data sets. 

Here, we circumvent this limitation with a twist: imagine first that we have at hand a sample $Y$ that is an independent copy of $X$. Then one could estimate $\para$ by an estimator $\wh{\para}(Y)$ computed from the sample $Y$. Therefore, 
\begin{align*}
U_{\beta(1-\delta)}(X,S(X);{\para}) &=  U_{\beta(1-\delta)}(X,S(X);\wh{\para}(Y))  +  \left(U_{\beta(1-\delta)}(X,S(X);{\para})  -U_{\beta(1-\delta)}(X,S(X);\wh{\para}(Y)) \right)\\
&\leq U_{\beta(1-\delta)}(X,S(X);\wh{\para}(Y))  + q_{2,\beta\delta}(\beta(1-\delta),S(X);\para),
\end{align*}
with probability larger than $1-\beta\delta$, where we denote
\begin{align*}
&q_{2,\gamma}(\beta,S(X);\para)\\
&= \min\left\{x\in \R \::\: \P_{Y\sim P_\para}\left(U_{\beta(1-\delta)}(X,S(X);{\para})  -U_{\beta(1-\delta)}(X,S(X);\wh{\para}(Y)) \leq x\:\big|\: X\right)\geq 1-\gamma\right\}.
\end{align*}
the $(1-\gamma)$-quantile of  the distribution of $U_{\beta(1-\delta)}(X,S(X);{\para})  -U_{\beta(1-\delta)}(X,S(X);\wh{\para}(Y))$ conditionally on $X$, when $Y\sim P_\Gamma$.
Similarly to above, the latter can be approximated via a bootstrapped quantity
 $
q_{2,\gamma}(\beta,S(X);\para)\approx q_{2,\gamma}(\beta,S(X);\wh{\para}(Y))$
which is the $\gamma$-quantile of the distribution of $U_{\beta(1-\delta)}(X,S(X);\wh{\para}(Y))  -U_{\beta(1-\delta)}(X,S(X);\wh{\para}(Y^*))$ conditionally on $X,Y$, when $Y^*\sim P_{\wh{\para}}$,
which itself is approximated by applying a Monte-Carlo approximation scheme, which gives rise to 
 $\tilde{q}^{(B)}_{2,\gamma}(\beta,S(X);\wh{\para}(Y))$. 
Now, since we do not have access to a different sample $Y$, we propose to use $X$ in place of $Y$, which leads  to the following new bootstrap bound (denoting, again, $\wh{\para}=\wh{\para}(X)$):
\begin{equation}\label{Qboot2}
\Ubootbis_{\beta,\delta}(X,S(X))=U_{\beta(1-\delta)}(X,S(X);\wh{\para})+ \tilde{q}^{(B)}_{2,\beta\delta}(\beta(1-\delta),S(X);\wh{\para}) ,
\end{equation}
Algorithm~\ref{algo:boot_bounds} gives full details about the computation of this bound.
 
Let us make the following comments on the second bootstrap bound:
\begin{itemize}
\item $\tilde{q}^{(B)}_{2,\beta\delta}(\beta(1-\delta),S(X);\wh{\para})$ does not depend on the full selection policy $S(\cdot):x\in \mathcal{X}\mapsto S(x)\subset\Nm$, but only on the set $S(X)$;
\item Whenever $\tilde{q}^{(B)}_{2,\beta\delta}(\beta(1-\delta),S(X);\wh{\para})$ is nonnegative 
 the obtained bound is more conservative than the plug-in bound.
However, while $\Ubootbis_{\beta,\delta}$ includes the variability in $\wh{\para}$, this bound ignores the variations in $S(X)$, so is in general less conservative than $\Uboot_{\beta,\delta}$. 
\end{itemize}

\subsection{Third bootstrap bound}\label{sec:boot3}

An elementary point is that the interval $I_\alpha(S(\cdot)) = [0,V_{\alpha}(S(\cdot);\para)]$ satisfies the unconditional coverage  \eqref{posthocbound3} when choosing
$$
q^{\textrm{naive}}_{\beta}(S(\cdot);{\para}) = \min\left\{x\in \R \::\: \P_\para\left(\FDP(\theta,S(X))  \leq x\right)\geq 1-\beta\right\}.
$$
{Note that this bound is ``unconditional'' and based on the distribution of $\FDP(\theta,S(X))$ when drawing $(\theta,X)\sim P_\para$. Hence, it uses the selection policy $S(\cdot)$.}  
This leads to choose $$\Unnaive_{\beta}(S(\cdot))=q^{\textrm{naive}}_{\beta}(S(\cdot);\wh{\para}).$$ {This bound thus relies on drawing independent couples $(\theta^*,X^*)$ from the distribution $P_{\wh{\para}}$. 
This is different from the above-mentioned bounds (bootstrap 1, bootstrap 2), which are based only on the marginal  $X^*$. 
}

As we will see in Section~\ref{sec:num}, using $\Unnaive_{\beta}$ is generally too conservative,
{ in particular if the true $\FDP(\theta,S(X))$ has large variability from a realization of $(\theta,X)$ to another, since it is based on an unconditional quantile of its distribution. To alleviate this effect, an improvement can be derived by using a proper re-centering by the plug-in bound $U_{\beta}(X,S(X);\wh{\para})$ given in \eqref{Qplugin}, that acts like a stabilization:
\begin{align}
V_{\beta}(X,S(\cdot);{\para}) &=\:U_{\beta}(X,S(X);\wh{\para}(X)) +q_{3,\beta}(\beta,S(\cdot);{\para})\label{Uuncond0}\\
q_{3,\beta}(\beta,S(\cdot);{\para})&=\min\left\{x\in \R \::\: \P_{(\theta,X)\sim P_\para}\left(\FDP(\theta
,S(X)) -U_{\beta}(X,S(X);\wh{\para}(X))   \leq x\right)\geq 1-\beta\right\}\label{equ-q3}
,
\end{align}
where $U_{\beta}(X,S(X);\para)$ is given by \eqref{equVnew} and \eqref{equVnew2}. 
{ By definition, $V_{\beta}(X,S(\cdot);{\para})-U_{\beta}(X,S(X);\wh{\para}(X)) $ is deterministic and  we have
\begin{align*}
&1-\beta\\
&\leq \P_{(\theta,X)\sim P_\para}\left(\FDP(\theta,S(X)) -U_{\beta}(X,S(X);\wh{\para}(X))   \leq V_{\beta}(X,S(\cdot);{\para})-U_{\beta}(X,S(X);\wh{\para}(X))\right)\\
&=\P_{(\theta,X)\sim P_\para}\left(\FDP(\theta,S(X))  \leq V_{\beta}(X,S(\cdot);{\para})\right).
\end{align*}
}
{This gives that $V_{\beta}(X,S(\cdot);\para) $ is a valid post hoc bound, which can be approximated by the bootstrap bound 
$$V_{\beta}(X,S(\cdot);\wh{\para})=U_{\beta}(X,S(X);\wh{\para}(X)) +q_{3,\beta}(\beta,S(\cdot);\wh{\para}).$$}
 {Given  \eqref{equ-q3}, this bound also relies on drawing the couple $(\theta^*,X^*)$ from the distribution $P_{\wh{\para}}$. Now, $q_{3,\beta}(\beta,S(\cdot);\wh{\para})$ is approximated by $\tilde{q}^{(B)}_{3,\beta}(\beta,S(\cdot);\wh{\para})$ via a Monte Carlo scheme  that generates $B$ times the couple $(\theta^*,X^*)$  according to the model with parameter $\wh{\para}$. This leads to the final bound:}
\begin{align}\label{equ:boot3}
\Unrecentred_{\beta}(X,S(\cdot))&=U_{\beta}(X,S(X);\wh{\para}) +\tilde{q}_{3,\beta}^{(B)}(\beta,S(\cdot);\wh{\para})
\end{align}
Algorithm~\ref{algo:boot_bounds} gives full details about the computation of this bound.

 The difference between $\Unrecentred_{\beta}(X,S(\cdot))$ and $\Unnaive_{\beta}(S(\cdot))$ is that our plug-in bound $U_{\beta}(X,S(X);\wh{\para})$ is used to recenter the FDP.  As a result, it ``stabilizes'' the bound $\Unnaive_{\beta}(S(\cdot))$: while $\Unnaive_{\beta}(S(\cdot))$ depends on $X$ only via $\wh{\para}$, the bound $\Unrecentred_{\beta}(X,S(\cdot))$ also depends on the plug-in bound $U_{\beta}(X,S(X);\wh{\para}(X))$.
Note that knowledge of the full selection policy $S(\cdot)$ is required
to compute both bounds $\Unnaive(S(\cdot))$ and $\Unrecentred_{\beta}(X,S(\cdot))$.

\section{Numerical experiments}
\label{sec:num}
\input{num_exp}

\section{Applications}
\label{sec:appli}
\input{appli}

\section{Discussion}
\subsection{Detecting close to singular scenarios}

{The HMM model becomes singular in the situation where $f_0=f_1$, or when the
transition matrix $A$ is of rank 1 (i.e. the coordinates of the configuration vector $\theta$ are drawn i.i.d. from a Bernoulli distribution), in which case the model is not identifiable:
the same data distribution for the observable $X$ can be obtained for several sets of parameters, in particular different joint distributions of $(\theta,X)$ so that the notion of
ground truth for the FDP is questionable.}

{The behavior of our approach in a situation close to singular was discussed
  in Section~\ref{sec:chall-model-assumpt}; in a truly singular situation, almost surely the estimator
$\wh{A}$ is not of rank 1, so that the estimation error (estimated at $\wh{\Gamma}$) will surely be underestimated by the bootstrap procedure.}

{It would be therefore in principle necessary to have a test for the singular case and stop
the procedure if that test is not rejected (using an agnostic multiple testing procedure would then
be more appropriate; see also the discussion on null estimation in the independent case,
Section~\ref{sec:relation}). While we do not cover the precise design of such a test in the present work,
a suitable heuristic procedure in practice can be to monitor if $\abs{\det \wh{A}}$ is too close to 0,
or $\wh{f}_1$ is too close to $f_0$. 

\subsection{Asymptotic consistency of plug-in}\label{sec:consistency}

{An important theoretical insight for FDR control methods using empirical Bayes-type approaches
  is that asymptotic consistency is usually granted for plug-in methods
  \citep{sun2009large}. In the case of nonparametric HMMs, consistent estimation with
quantitative guarantees for the HMM parameters and the local FDR have been obtained recently, see discussion in Section~\ref{sec:relation}. In the setting considered here, since we
are considering in principle arbitrary policies and therefore selection sets $S(X)$, it seems
we would need a stronger convergence property for parameter estimation, e.g. in the sense of total variation distance for the full joint distribution of $(\theta,X)$. If such a convergence
holds, we can guarantee the asymptotic of the plug-in approach for FDP bounds (see
Appendix~\ref{app:const}.)}

{However, obtaining such a consistency result for convergence of
  the full estimated joint distribution seems a tall order, since the asymptotics would be in
  the increasing size of the observation space, which will also make TV distance larger. (It might
  be more plausible for theory to assume that the parameter is estimated using an observation of larger size than the one it is used to perform multiple testing on.) Furthermore, our experiments show
  that the simple plug-in approach is generally unsatisfactory in practice and that the proposed
bootstrap-based bounds are more appropriate.}

\subsection{Bayesian post hoc versus frequentist post hoc}\label{sec:discuss}

{
  The posterior (oracle) bound~\eqref{posthocbound4} can
  accommodate absolutely any selection policy $S(\cdot)$. Thus it seems that the bound holds
  in fact for any selected set $R=S(X)$. This can appear surprising at first, since in
  a frequentist setting, uniform agnostic post-hoc bounds (holding for all subsets $R$ and over all
  latent parameter configurations $\theta$) generally have a price for complexity
  (e.g. a union bound over a reference family of candidate rejection sets), while the
  Bayesian approach seems to offer a free lunch in that regard.}

A key to understand this apparent conundrum is to insist that
  \begin{enumerate}
  \item  strictly speaking, the guarantees only hold if the assumed structural model (prior) on the latent variable
    $\theta$ is correct. In Bayesian terminology, if $\P_\para$ is considered a prior, then
    the proposed (oracle) bounds are credible sets on the posterior FDP.
  \item   the posterior bound~\eqref{posthocbound4} is {\em not} a uniform statement over all possible rejection sets,
  but should be interpreted as conditioning with respect to $X$ of an arbitrary but given
  selection policy $S(\cdot)$ that must {\em only depend on the observation $X$}. 
  \end{enumerate}
 This excludes to use any form of
  ``insider'' or ``leaked'' information on the latent variable configuration to be used for to
  determine the selected set, such that:
  \begin{itemize}
  \item Ancillary statistics conveying additional information;
  \item ``Expert knowledge'' that would not be incorporated in the prior, e.g. an expert
    communicating that a certain region has a higher chance of containing false hypotheses,
    or that false hypothesis configurations should have a certain shape, while the prior does
    not distinguish a priori such configurations. This would be akin to using a prior that
    we know from the expert information is wrong (we can also interpret
    the expert knowledge as ``insider information'' on the configuration under the original prior).
  \end{itemize}
As discussed in Section~\ref{sec:chall-model-assumpt}, violation of these model assumptions
  can result in the oracle bound to be incorrect. 
  From a Bayesian point of view, the prior structural distribution assumption on $\theta$ should
  reflect the entirety of the available prior information, and not be used merely
  as a convenient default. 
  
Connected to this, it would be interesting to study whether ``frequentist Bayes'' approaches  based on posterior concentration to the true latent parameter (see, e.g., \citealp{CR2020}, for a recent review) would be applicable here. 
    While recent progress has been achieved to analyze the FDR from this perspective \citep{CR2020,abraham2021multiple,abraham2021empirical},
    the same question relative to post hoc bounds has not been explored yet up to our knowledge.

  \section*{Acknowledgements}
  
This work has been supported by ANR-16-CE40-0019 (SansSouci), ANR-17-CE40-0001 (BASICS), ANR-19-CHIA-0021-01 (BiSCottE), the UPSaclay Excellency Chair REC-2019-044, and by the GDR ISIS through the "projets exploratoires" program (project TASTY).

\section*{Appendix}
\appendix

\addcontentsline{toc}{section}{Appendix}

\section{Posterior distribution}\label{sec:post}

We provide here the details for the quantities involved in the posterior distribution, see Proposition~\ref{prop:thetacondX}.
{We recall that $\pi=(\pi_0,\pi_1)$ denotes the marginal stationary distribution of
the underlying Markov Chain.} Let for $i\in \Nm$ and $q\in\{0,1\}$,
\begin{align}
\alpha_i(q) &= \P_\para(X_{1:i}, \theta_i =q);\nonumber\\
\beta_i(q) &= \P_\para(X_{(i+1):m}\:|\: \theta_i = q);\nonumber\\
\bell_{i, q}(\para)&= \P_\para(\theta_i = q\:|\: X_{1:m})=\frac{\alpha_{i}(q)\beta_i(q)} {\sum_{k \in\{0,1\}} \alpha_{i}(k)\beta_i(k)},
\label{equlti}
\end{align}
with the common notation for which $\P_\para(X_{1:i}, \theta_i =q)$ denotes the density of $(X_{1:i}, \theta_i)$ taken at point $(X_{1:i},q)$ and $\P_\para(X_{(i+1):m}\:|\: \theta_i = q)$ denotes the density of $X_{(i+1):m}$ conditionally on  $\theta_i = q$ taken at point $X_{(i+1):m}$.
These quantities can be obtained through the standard forward-backward algorithm, that is,
\begin{align}
\alpha_1(q) = \pi_{q} f_q(X_1)  ,\:\:\alpha_{i+1}(q) = f_{q}(X_{i+1}) \sum_{\l\in\{0,1\}} a_{\l,q} \alpha_{i}(\l)  , \:\:\:1\leq i \leq m-1 \label{eq:fwbc_al}\\
\beta_m(q)= 1,\:\: \beta_{i-1}(q)= \sum_{\l\in\{0,1\}} f_{\l}(X_{i}) a_{q,\l} \beta_{i}(\l) ,\:\:\: 2\leq i\leq m. \label{eq:fwbc_bet}
\end{align}
Further, we let for $i\in \{2,\dots,m\}$ and $q,q'\in\{0,1\}$,
\begin{align}
\bell_{i, q,q'}(\para)&= \P_\para(\theta_i = q' , \theta_{i -1} = q | X_{1:m})
=\frac{\beta_i(q')\alpha_{i-1}(q)f_{q'}(X_i){ a}_{q,q'}} { \sum_{k \in\{0,1\}} \beta_i(k)\alpha_{i}(k)};
\label{equltij}\\
%
  \Post_{i,q,q'}(\para) &= { \P_\para(\theta_i=q'|\theta_{i-1}=q,X_{1:m})}
                          = \frac{\bell_{i, q,q'}(\para)}{\bell_{i-1, q}(\para)}
=\frac{\beta_i(q')f_{q'}(X_i){ a}_{q,q'}}{ \beta_{i-1}(q)};
\label{posterior}\\
\Post_i(\para) &= \left(
\begin{matrix}
\Post_{i,0,0}(\para)  & \Post_{i,0,1}(\para) \\
\Post_{i,1,0}(\para) & \Post_{i,1,1}(\para) \\
\end{matrix}
\right).\label{equATX}
\end{align}

\section{Details of the implementation}
\subsection{Details on Algorithm~\ref{algo:EM}}

The initialization $\wh{\para}^{(0)}$ of $\wh{\para}$ in Algorithm~\ref{algo:EM} is done as follows: $\wh{A}^{(0)}$ is computed first by estimating the stationary distribution, that is, $(\pi_0,1-\pi_0)$, by using a standard Storey estimator $\wh{\pi}^{(0)}_0$ ($\lambda=0.8$) \cite{Storey2002direct}. Then, the parameter $\wh{A}^{(0)}_{1,1}$ is taken uniformly over the possible range for this parameter, intersected with $[0.6,1]$ to ensure that the null class is predominant.
 On the other hand, $\wh{f}^{(0)}_1= (\wh{f}-\wh{\pi}^{(0)}_0 f_0)/(1-\wh{\pi}^{(0)}_0)$, where $\wh{f}$ is a Kernel-based estimation of the density of the measurements. 
The algorithm is stopped if the distance between to successive iteration of $\wh{\para}$ (measured in infinite norm, restricted on the observations for the densities) is below $10^{-4}$.

\subsection{Details on Algorithm~\ref{algo:Bkl} }

In practice for a large $s$ Algorithm~\ref{algo:Bkl} can be time consuming because it requires the estimation of 2 $s\times s$ matrices. 
To speed up the algorithm we notice that it is not necessary to compute the matrices $B_{k,\l,0}$ and $B_{k,\l,1}$ for all $\l \in\{1,\dots, s\}$ to get $U_\beta(X,S(X),\para)$. Indeed, we can stop for $u$ such that $B_{k,u,1} + B_{k,u,0} \geq 1 - \beta$: as shown in Proposition~\ref{propnewbound} $U_\beta(X,S(X),\para) = u$.

\subsection{Details on Algorithm~\ref{algo:boot_bounds}}
\label{sec:app_impl_boot}
\paragraph{Drawing $\theta^{*(b)}$ and $X^{*(b)}$ from $\wh{\para}$}

\begin{enumerate}
\item Draw a Markov chain $\theta^{*(b)}$ of size $m$ with transition matrix $\wh{A}$, and initial distribution the stationary distribution, namely $(\frac{\wh{a}_{1,0}}{\wh{a}_{0,1}+ \wh{a}_{1,1}}, 1 - \frac{\wh{a}_{1,0}}{\wh{a}_{0,1}+ \wh{a}_{1,1}})$.
\item For $1\leq i\leq m$ draw independently $X_i^{*(b)}$ as (case $f_0$ known):
$$
X_i^{*(b)}\sim \left\{\begin{array}{cc}
f_0 & \mbox{ if $\theta_i^{*(b)}=0$};\\
\wh{f}_1 & \mbox{ if $\theta_i^{*(b)}=1$};
\end{array}\right.
$$
Drawing from $\wh{f}_1$ can be done easily by noting that $\wh{f}_1$ is the density of a mixture of Gaussian distributions:
$$
\wh{f}_1 = \sum_{i=1}^m w_i\: \mathcal{N}(X_i,h^2), \:\:\: w_i=\frac{\bell_{i, 1}(X;\wh{\para}) }{\sum_{j=1}^m \bell_{j, 1}(X;\wh{\para}) }, \:\:1\leq i\leq m.
$$
If $f_0$ is unknown, the only difference is that $X_i^{*(b)}\sim \wh{f}_0$ when $\theta_i^{*(b)}=0$. Drawing from $\wh{f}_0$ is made similarly to the case of $\wh{f}_1$.
\end{enumerate}

\paragraph{Empirical quantiles}
To obtain the empirical quantiles $\tilde{q}_{i,\gamma}$  of $D_i = (D_i^{(1)},\dots,D_i^{(B)})$ we start by ordering $D_i$. 
More precisely let $b_1,\dots,b_B$ such that 
$
D_i^{(b_1)}\leq D_i^{(b_2)} \leq \dots \leq D_i^{(b_B)}
$
then we define
$\tilde{q}_{i,\gamma} = D_i^{(b_j)}$, where
 $j$ is the smallest integer larger or equal to $\gamma B$.

\section{Computation of the  lower bounds}

\label{sec:lowerbounds}

Expressions for lower bounds can be obtained similarly to those of upper bounds. We provide them in this section for completeness.
\paragraph{Oracle and plug-in lower bounds.}
According to \eqref{posteriorlowerbound}, we have (using the notation of Section~\ref{sec:oraclebound}, e.g., $R=S(X)$, $s=|S(X)|$):
\begin{equation*}
L_{\beta}(X,R;\para)=s^{-1} \max\left\{n\in \{0,\dots,m\}\::\: \P_\para\left(\sum_{t=1}^s (1-\theta^R_{t})\geq n\:\bigg|\: X\right)\geq 1-\beta
 \right\}.
\end{equation*}

Similarly to Proposition~\ref{propnewbound}   the quantity $L_{\beta}(X,S(X);\para)$ can be computed as 
\begin{equation*}
L_{\beta}(X,R;\para)=s^{-1} \max\left\{n\in \{0,\dots,m\}\::\:B_{s,n-1,0} +B_{s,n-1,1} \leq \beta \right\}.
\end{equation*}
Adding the convention $B_{s,-1,0} =B_{s,-1,1}=0$.
The plug-in lower bound is given by $L_{\beta}(X,S(X);\wh{\para})$.

\paragraph{First  bootstrap lower bound.}
 
Proceeding as for the upper bound (see Section~\ref{sec:boot1}), we obtain
\begin{align*}
\P_\para( \FDP(\theta,S(X))\geq L_{\beta(1-\delta)}(X,S(X);\para)\:|\: X)&\geq 1-\beta(1-\delta)\\
\P_\para(L_{\beta(1-\delta)}(X,S(X);{\para})-L_{\beta(1-\delta)}(X,S(X);\wh{\para})\geq q^{\l}_{1,\beta\delta}(\beta(1-\delta),S(\cdot);\para))&\geq 1-\beta\delta.
\end{align*}
by letting
$$
q^{\l}_{1,\gamma}(\beta,S(\cdot);\para)) = \max \{ x \in\mathbb{R} \::\:\;\; \P_\para(L_\beta(X,S(X),\para) -L_\beta(X,S(X),\wh{\para}) \geq x) \geq 1 -\gamma\}.
$$
This gives
 \begin{equation}\label{equ:covboot2}
 \P_\para(\FDP(\theta,S(X))\geq L_{\beta(1-\delta)}(X,S(X);\wh{\para})+ q^{\l}_{1,\beta\delta}(\beta(1-\delta),S(\cdot);\para))\geq 1-\beta.
 \end{equation}
and leads to the boot1 lower bound:
$$
L^{\textrm{boot1}}_{\beta,\delta}(X,S(\cdot)) = L_{\beta(1-\delta)}(X,S(X);\wh{\para})+ \tilde{q}^{\l, (B)}_{1,\beta\delta}(\beta(1-\delta),S(\cdot);\widehat{\para}),$$
where  $\tilde{q}^{\l, (B)}_{1,\beta\delta}(\beta(1-\delta),S(\cdot);\widehat{\para})$ is the Monte-Carlo approximation of the bootstrap quantile ${q}^{\l}_{1,\beta\delta}(\beta(1-\delta),S(\cdot);\widehat{\para})$, itself being an approximation of ${q}^{\l}_{1,\beta\delta}(\beta(1-\delta),S(\cdot);{\para})$. In practice, $\tilde{q}^{\l, (B)}_{1,\beta\delta}(\beta(1-\delta),S(\cdot);\widehat{\para})$ can be derived as $E_1^{(b_i)}$ with $i = \lfloor B \beta\delta\rfloor + 1$ where
$E_1^{(b_1)} \leq E_1^{(b_2)}\leq \dots \leq E_1^{(b_B)}$
with 
$$
E_1^{(b)} = L_{\beta(1-\delta)}(X^{*(b)},S(X^{*(b)}),\wh{\para}) -L_{\beta(1-\delta)}(X^{*(b)},S(X^{*(b)}),\wh{\para}^{*(b)}), 1\leq b\leq B.
$$

\paragraph{Second bootstrap lower  bound.} 
The same heuristic as for $\Ubootbis$ (see Section~\ref{sec:boot2}) gives:
$$
L^{\textrm{boot2}}_{\beta,\delta}(X,S(X)) = L_{\beta(1-\delta)}(X,S(X);\wh{\para})+ \tilde{q}^{\l, (B)}_{2,\beta\delta}(\beta(1-\delta),S(X);\widehat{\para}),$$
where  $\tilde{q}^{\l, (B)}_{2,\beta\delta}(\beta(1-\delta),S(X);\widehat{\para})=E_2^{(b_i)}$ with $i = \lfloor B \beta\delta\rfloor + 1$ where
$E_2^{(b_1)} \leq E_2^{(b_2)}\leq \dots \leq E_2^{(b_B)}$
with 
$$
E_2^{(b)} = L_{\beta(1-\delta)}(X,S(X),\wh{\para}) -L_{\beta(1-\delta)}(X,S(X),\wh{\para}^{*(b)}), 1\leq b\leq B.
$$

\paragraph{Third lower bootstrap bound.}
Using similar arguments as for $\Unrecentred$ (see Section~\ref{sec:boot3}), we obtain the bound 
$$
L^{\textrm{boot3}}_{\beta}(X,S(\cdot)) = L_\beta(X,S(X),\wh{\para}) + \tilde{q}_{3,\beta}^{\l,(B)}(\beta,S(\cdot);\wh{\para}),
$$
where  $\tilde{q}^{\l, (B)}_{3,\beta}(\beta,S(X);\widehat{\para})=E_3^{(b_i)}$ with $i = \lfloor B \beta\rfloor + 1$ where
$E_3^{(b_1)} \leq E_3^{(b_2)}\leq \dots \leq E_3^{(b_B)}$
with 
$$
E_3^{(b)} = \FDP(X^{*(b)},S(X^{*(b)})) -L_\beta(X^{*(b)},S(X^{*(b)}),\wh{\para}^{*(b)}) , 1\leq b\leq B.
$$

\section{Estimation of $f_0$}\label{estimf0}

When $f_0$ is unknown, we should estimate it along with $A,f_1$ in Algorithm~\ref{algo:EM}. 
The following algorithm builds an estimator $\wh{f}_0$ in the same spirit as $\wh{f}_1$ \eqref{eq:estf1}, by replacing the $\bell_{i, 1}$ by $\bell_{i, 0}$, that is, 
\begin{equation}\label{eq:estf0}
\wh{f}^{(t)}_0(x) = \sum_{i=1}^m \bell_{i, 0}(\wh{\para}^{(t-1)})\frac{K((x-X_i)/h)}{h} / \sum_{i=1}^m \bell_{i, 0}(\wh{\para}^{(t-1)}).
\end{equation} 

\begin{algorithm}[H]
 \KwData{$X=(X_1,\dots,X_m)$}
 \KwResult{$\wh{\para}=(\wh{A},\wh{f}_0,\wh{f}_1)$ estimator of $\para=(A,f_0, f_1)$}
 \begin{itemize}
 \item
  Initialization: first guess $\wh{\para}^{(0)}=(\wh{A}^{(0)},\wh{f}_0^{(0)},\wh{f}_1^{(0)})$ of $\para$
 \item
 Loop: at step $t\geq 1$, given $\wh{\para}^{(t-1)}$, do
 \begin{itemize}
 \item
 Compute 
\begin{enumerate}
\item $\alpha_i(q)=\alpha_i(q;\wh{\para}^{(t-1)})$ and $\beta_i(q)=\beta_i(q;\wh{\para}^{(t-1)})$, $1\leq i\leq m$, $q \in \{0,1\}$, using the forward-backward algorithm \eqref{eq:fwbc_al}, \eqref{eq:fwbc_bet};
\item $\bell_{i, q}(X;\wh{\para}^{(t-1)}) $, $1\leq i\leq m$, $q \in \{0,1\}$, 
by using \eqref{equlti};
\item $\bell_{i, q,q'}(X;\wh{\para}^{(t-1)}) $, $1\leq i\leq m$, $q,q' \in \{0,1\}$, 
by using \eqref{equltij}.
\end{enumerate}
 \item
Update
\begin{enumerate}
\item $\wh{a}^{(t )}_{q,q'} = \sum_{i = 1}^{m-1}  \bell_{i, q,q'}(\wh{\para}^{(t-1)}) /  \sum_{i = 1}^{m-1}  \bell_{i, q}(\wh{\para}^{(t-1)})$ , for $ 0 \leq q,q' \leq 1$;
\item 
$
\wh{A}^{(t )}  =\left( \begin{matrix}
\wh{a}^{(t )}_{0,0} & \wh{a}^{(t )}_{0,1} \\
\wh{a}^{(t )}_{1,0} & \wh{a}^{(t )}_{1,1}
\end{matrix} \right);
$
\item $\wh{f}_0^{(t )}$ using \eqref{eq:estf0};
\item $\wh{f}_1^{(t )}$ using \eqref{eq:estf1}; 
\item $\wh{\para}^{(t )} = (\wh{A}^{(t )}, \wh{f}_0^{(t )},\wh{f}_1^{(t )})$.
\end{enumerate}
 \item
Stop the loop if convergence, that is, $\wh{\para}^{(t )}$ close enough to $\wh{\para}^{(t -1)}$. 
 \end{itemize}
 \item Return $\wh{\para}=\wh{\para}^{(t)} $
 \end{itemize}
 \caption{\label{algo:EMf0unknown} EM-type algorithm to derive $\wh{\para}=(\wh{A},\wh{f}_0,\wh{f}_1)$\:\: (with $f_0$ unknown).
 }
\end{algorithm}

Note that, at the end of Algorithm~\ref{algo:EMf0unknown}, the user can define the ``null'' state according to their preference. For instance:
\begin{itemize}
\item define the "null" state as the predominant one, that is, the most probable one according to the stationary distribution of $\wh{A}$;
\item define the "null" state according to the density among $\{\wh{f}_0, \wh{f}_1\}$ whose mean is closer to $0$.
\end{itemize}

\marie{When $f_0$ is estimated, we obtain as a by-product empirical $p$-values, denoted by $\wh{p}$: these are based on the estimated empirical cumulative function under $\cH_0$, namely $\wh{p}_i =2(\min(1 -\wh{F}_0(t_i),\wh{F}_0(t_i))$, where
$$
\wh{F}_0(x) = \frac{\displaystyle \sum_{x_i <x}\bell_{i,0}(\wh{\para}) }{\displaystyle \sum^m_{i =1}\bell_{i,0}(\wh{\para})}.
$$ 
 .}
 
\section{Additional numerical experiments}
\label{sec:addit-numer-exper}
In this section we present additional numerical experiments. We have modified the simulation described Section~\ref{sec:num} by changing either the model parameters $\wh{\para}$ or the selection policies $S(\cdot)$. In all these additional experiments, the number of hypotheses is $m=3200$, the number of runs is 300  and the risk is $\beta = 10 \%$.

\subsection{Invalid selection policies}
\label{sec:inval-select-polic}

The selection policies used in Figure~\eqref{fig:knowledge} produce invalid bounds (even the oracle bound) as they depend of some knowledge of $\cH_0$, the set of true null hypotheses.
\begin{figure}[!htp]
\includegraphics[width = \textwidth]{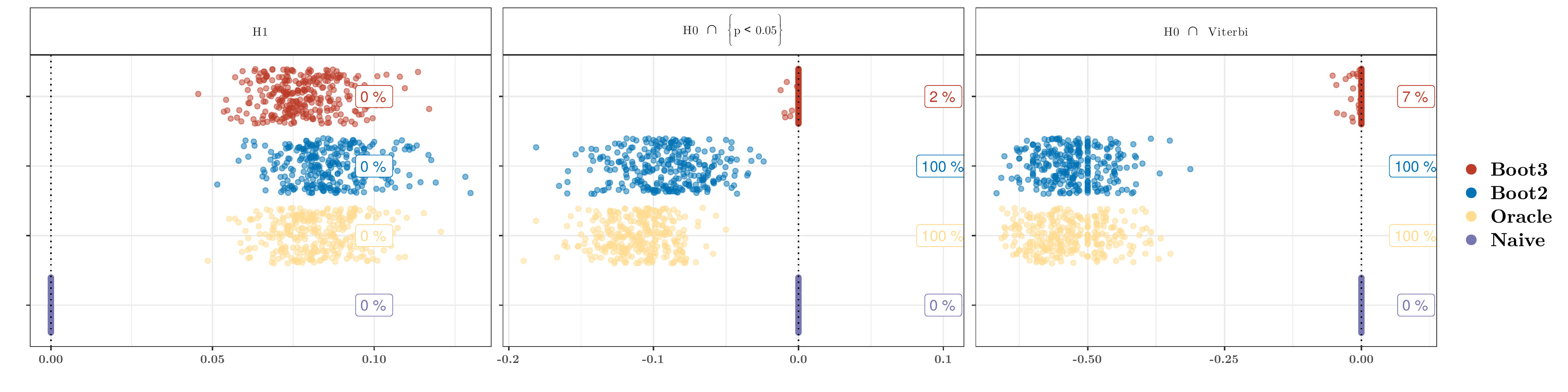}
\caption{Plot similar to Figure~\ref{fig:highdet} (A) (same model and simulation parameters) but when the selection policy is of the form $S(X,\cH_0)$, that is,  depends on some knowledge of $\cH_0$.
Recall that the targeted level $\beta$ is $10\%$.
}
\label{fig:knowledge}
\end{figure}

\subsection{Independent states or small determinant}

In Figure \ref{fig:diffdet}, we set $a_{0,0} = 0.95$, and the value of  $a_{1,1}$ is modified in order to achieve the desired value of $\det(A)$. 
\begin{figure}[ht]
\includegraphics[width = \textwidth]{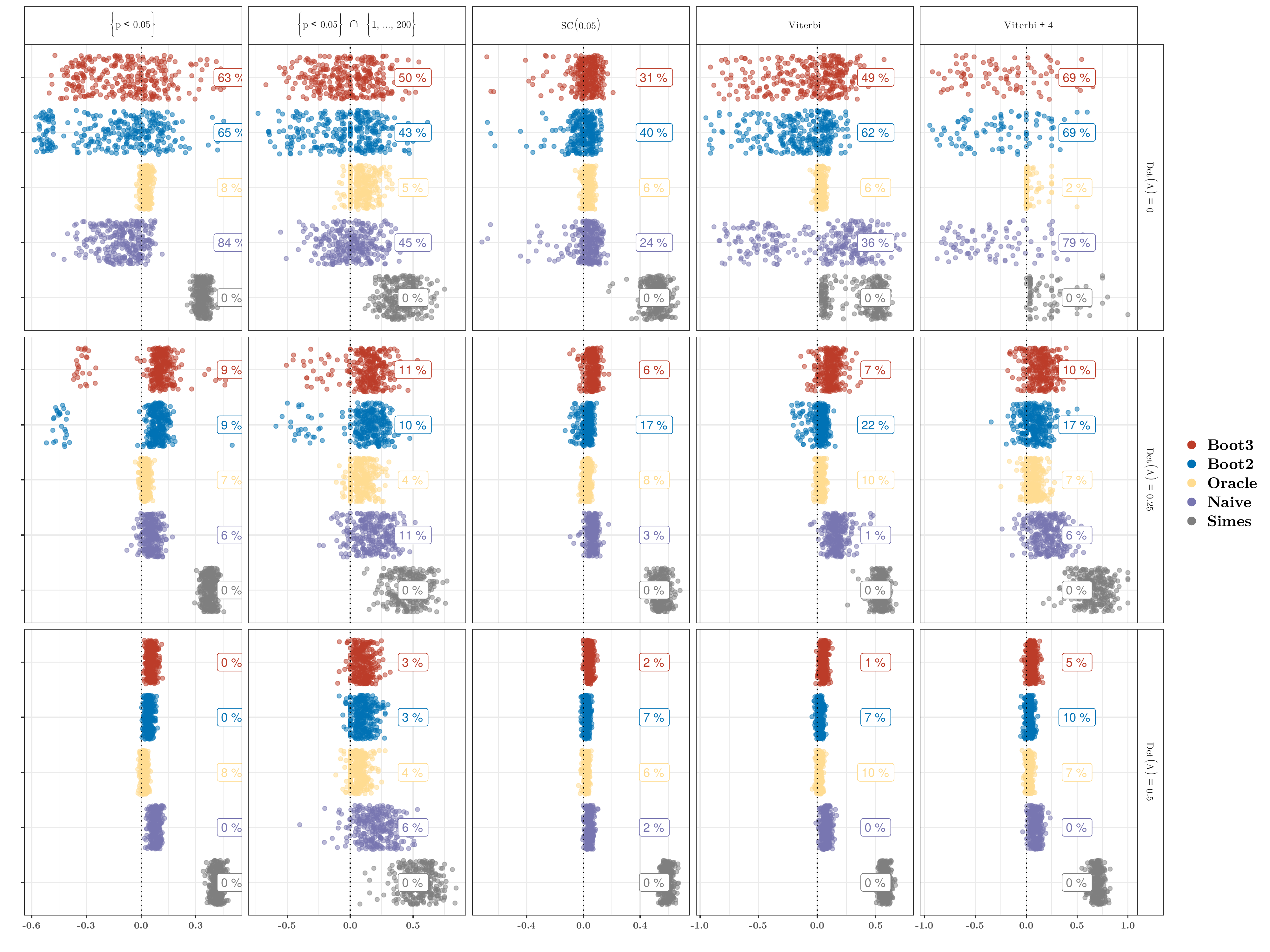}
\caption{Plot similar to Figure~\ref{fig:highdet} (A) (same densities $f_0$ and $f_1$, and same simulation parameters), for different model parameters making $\det(A)$ small (rows).
Recall that the targeted level $\beta$ is $10\%$.}
\label{fig:diffdet}
\end{figure}

\subsection{Unknown $f_0$}
\label{sec:suppl-unknown-f0}
Figure~\ref{fig:highdet_uf0} presents the results of numerical experiments in which $f_0$ is unknown and has to be estimated from the data as well.
 
 \begin{figure}[!htp]
 \begin{tabular}{cc}
 \includegraphics[width= \textwidth]{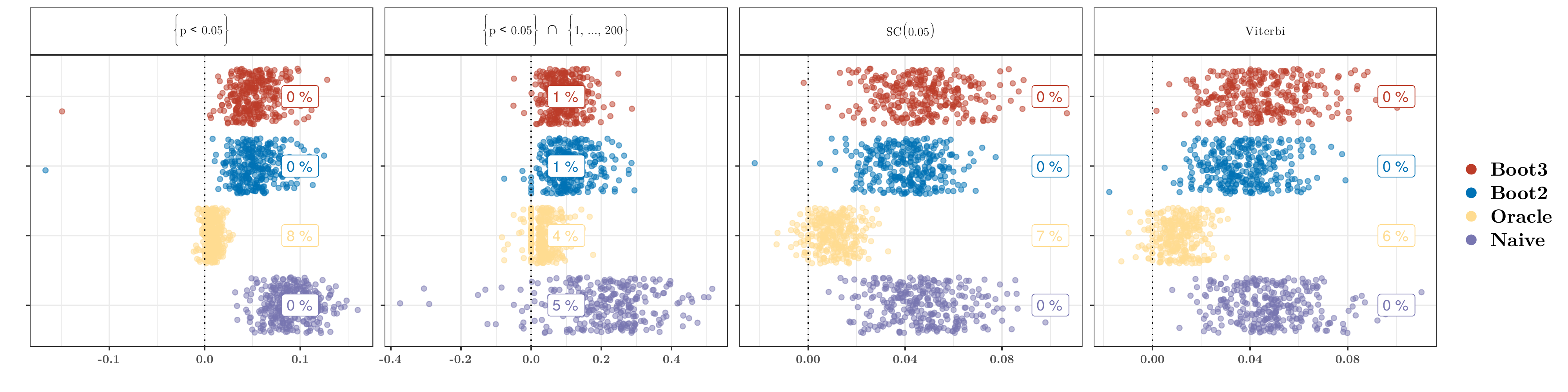} \\
 \includegraphics[width=\textwidth]{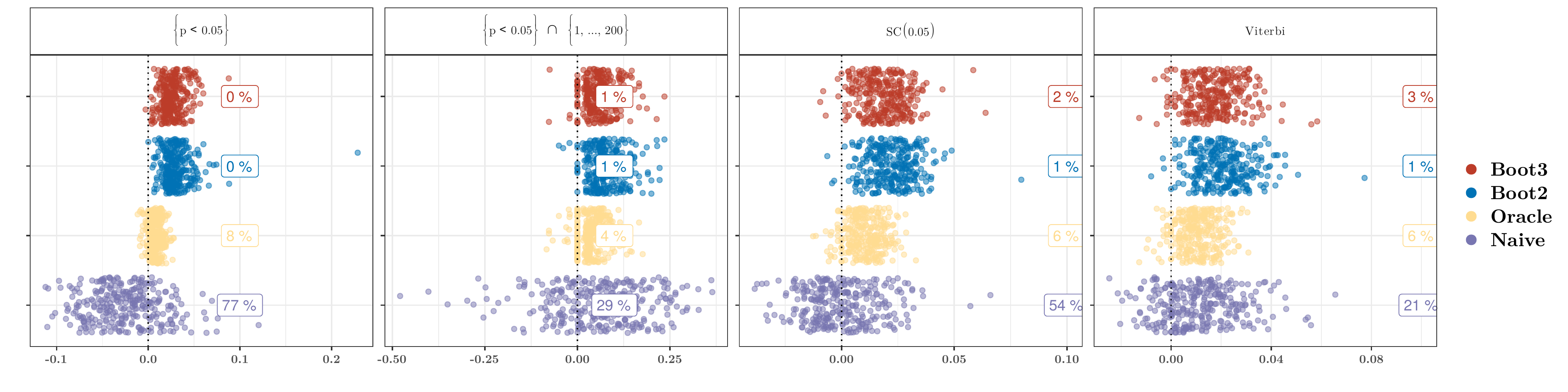}
 \end{tabular}
 \caption{Plot similar to Figure~\ref{fig:highdet} (A) (same model  and simulation parameters) when $f_0$ is unknown, that is, when the bounds also use a $f_0$ estimator, see Algorithm~\ref{algo:EMf0unknown}. Top panel:  initialization with the true $f_0$; Bottom panel: initialization using local  FDR algorithm ~\citep{efron2004large}}
 \label{fig:highdet_uf0}
 \end{figure} 
 
\subsection{Semi-simulated copy-number data}
\label{sec:suppl-plots-jointseg}
Figure~\ref{fig:jointseg-regions} illustrates the  semi-simulated data set analyzed in Section \ref{sec:simu-jointseg}.
\begin{figure}[!htp]
  \centering
  \includegraphics[width=0.95\textwidth]{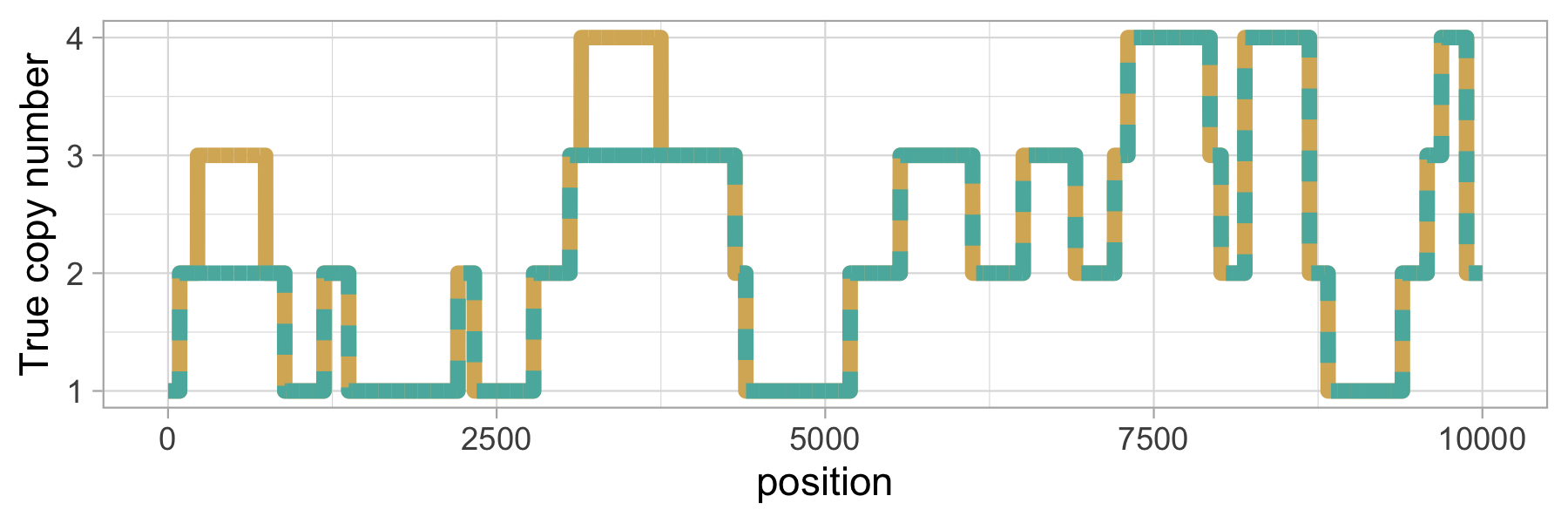}
  \includegraphics[width=0.95\textwidth]{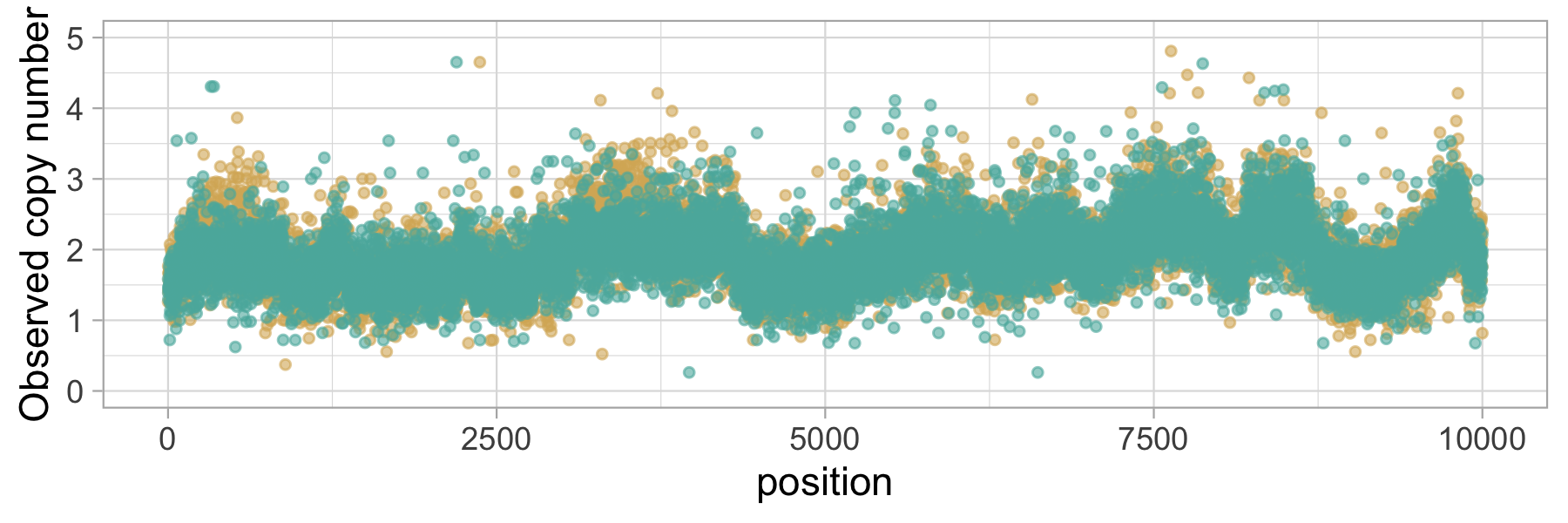}
  \includegraphics[width=0.95\textwidth]{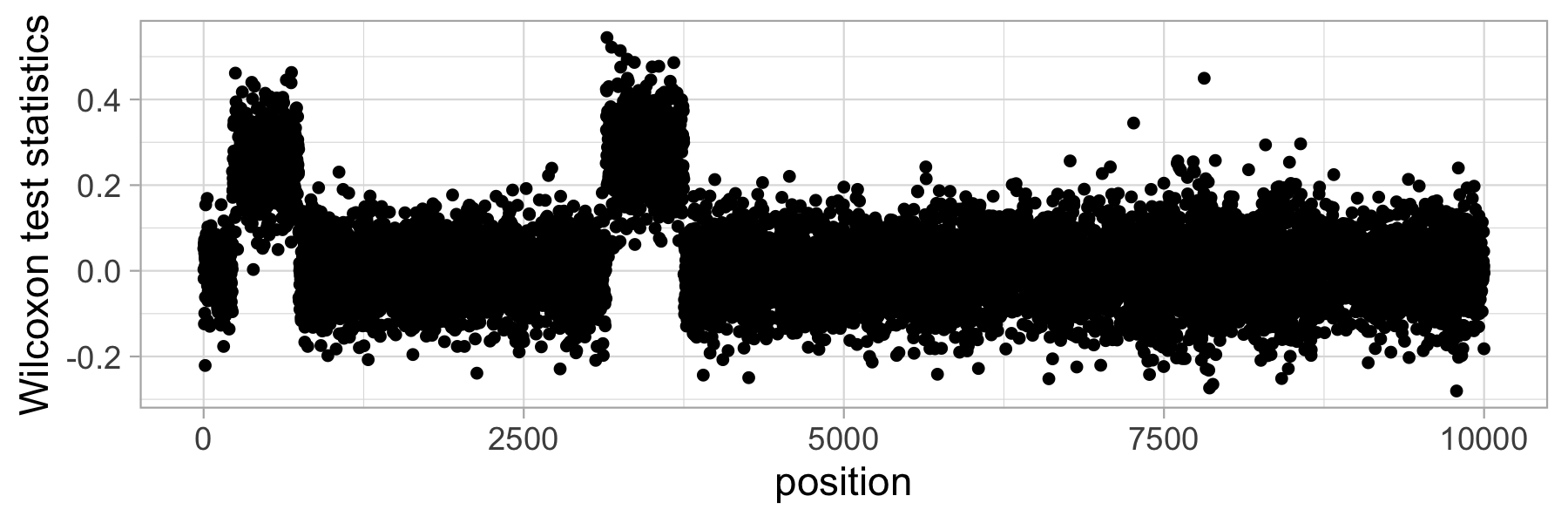}
  \caption{Illustration of the semi-simulated data set analyzed in Section \ref{sec:simu-jointseg}. Top: true CN regions; middle: CN signal for one sample for group 1 and one sample for group 2; bottom: Wilcoxon test statistics for the comparison of $n_1=50$ samples from group 1 and $n_2=50$ samples from group 2. Here the proportion of tumor cells is set to $70\%$.}
  \label{fig:jointseg-regions}
\end{figure}

\subsection{Power in the semi-simulated case}
\label{sec:powerrealist}
Figure~\ref{fig:reali-power} displays the power of the different bounds in the semi-simulated data set analyzed in Section \ref{sec:simu-jointseg}.
\begin{figure}[!htp]
\includegraphics[width=\textwidth]{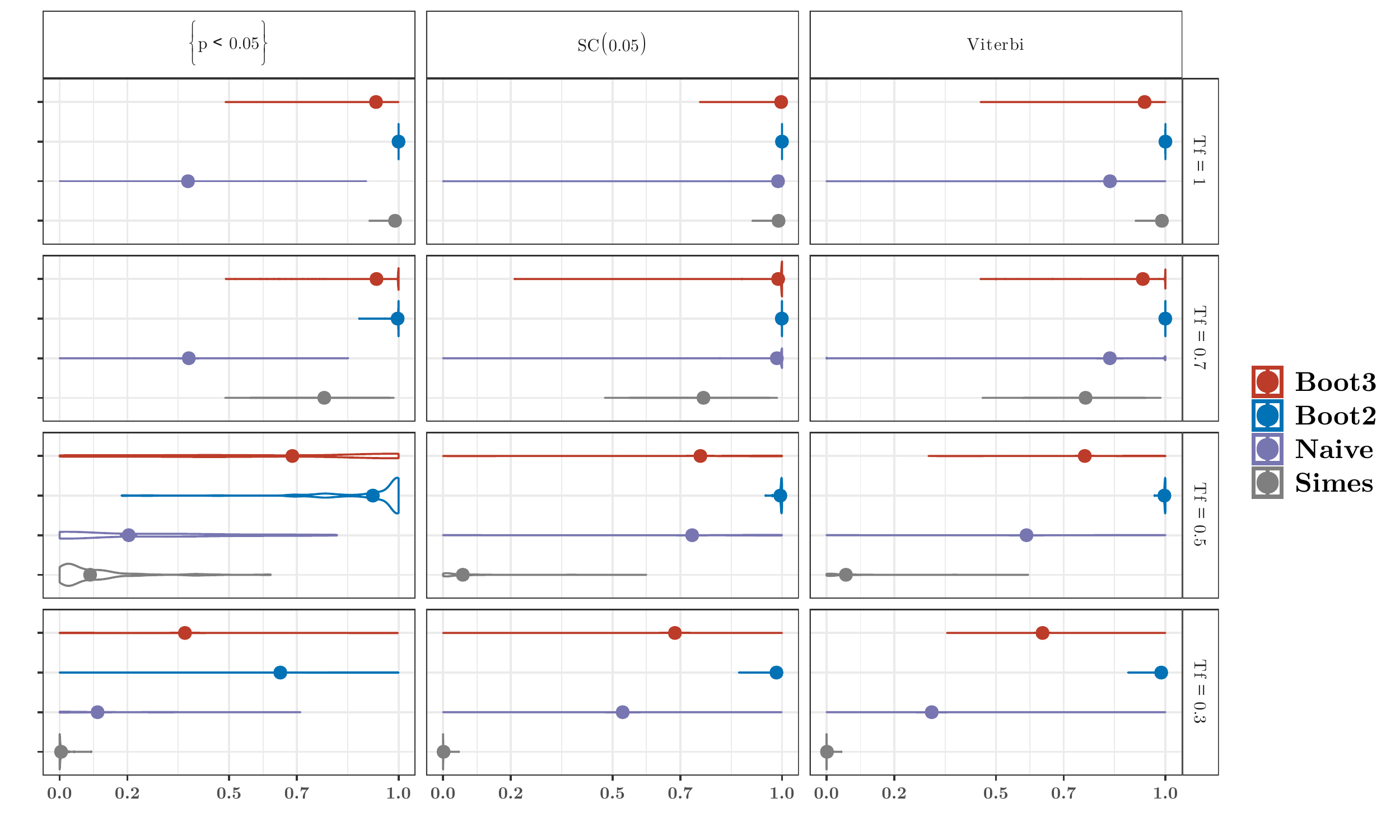}
\caption{
Summary of the power for the semi-simulated data set analyzed in Section \ref{sec:simu-jointseg}.}
\label{fig:reali-power}
\end{figure}
\section{Towards plug-in consistency}
\label{app:const}

The aim of this section is to provide sufficient conditions in order to ensure that the plug-in bound is asymptotically valid, as $m$ tends to infinity. This supports the discussion made in Section~\ref{sec:consistency}.

\begin{lemma}
Assume that $\wh{\para}$ is an estimator of $\para$ such that 
\begin{align}\label{condconsistency}
\forall \para,\:\:  d_{tv}\left( \mathcal{D}_{\wh{\para}}\left(\theta\:\big|\: X\right),\mathcal{D}_{{\para}}\left(\theta\:\big|\: X\right)  \right) \mbox{ converges in probability to $0$ under $\para$},
\end{align}
where $\mathcal{D}_{{\para}}\left(\theta\:\big|\: X\right)$ denotes the conditional distribution of $\theta$ conditionally on $X$ under $P_\para$ and $d_{tv}$ denotes the total variation distance.
Then the plug-in bound $\UPI_\beta(X,S(X))$ \eqref{Qplugin}
 satisfies that
$$\forall \para,\:\:\liminf_m\left\{ \P_\para\left(\FDP(\theta,S(X)) \leq \UPI_\beta(X,S(X))\right)\right\}\geq 1-\beta.$$ 
\end{lemma}

\begin{proof}
Since $\UPI_\beta(X,S(X))=U_{\beta}(X,S(X);\wh{\para})$, we have point-wise in $X$, for all $\para$, 
\begin{align*}
&\P_\para\left(\FDP(\theta,S(X))\leq  \UPI_\beta(X,S(X))\:\bigg|\: X \right)\\
=\:& \P_\para\left(\FDP(\theta,S(X))\leq  U_{\beta}(X,S(X);\wh{\para})\:\bigg|\: X \right)\\
\geq\:& \P_{\wh{\para}}\left(\FDP(\theta,S(X))\leq  U_{\beta}(X,S(X);\wh{\para})\:\bigg|\: X \right)\\
& -\sup_{n\in [0,m]}\left\{\P_{\para}\left(\sum_{i\in S(X)}(1-\theta_i)\leq  n\:\bigg|\: X\right)-\P_{\wh{\para}}\left(\sum_{i\in S(X)}(1-\theta_i)\leq  n\:\bigg|\: X\right)\right\}\\
\geq\:& 1-\beta - d_{tv}\left( \mathcal{D}_{\wh{\para}}\left(\theta\:\big|\: X\right),\mathcal{D}_{{\para}}\left(\theta\:\big|\: X\right)  \right),
\end{align*}
by definition \eqref{equVnew} of the functional $U_{\beta}$.
This entails the result by integrating over $X$.
\end{proof}

\addcontentsline{toc}{section}{References}
\bibliographystyle{habbrvnat}
\bibliography{biblio}

\end{document}

%% file: num_exp.tex
This section summarizes numerical experiments performed in order to assess the quality of the bounds introduced in Section~\ref{sec:ourposthocbounds}. 

\paragraph{Post hoc bounds.} These bounds will be compared to the Simes post hoc bound, a baseline competitor which does not take the latent structure of the model into account.  The Simes post hoc bound has been introduced by \cite{goeman2011multiple}; it provides an upper bound on the false positives, but no corresponding lower bound. It is valid as soon as the Simes inequality holds, which is the case here because the $p$-values are independent conditionally on $\theta$.
 Following \cite{blanchard2020post}, the Simes post hoc bound can be formulated as:
\begin{align}
U^{\textrm{Simes}}(S(X)) &= |S(X)|^{-1}\min_{1\leq k \leq m}\left\{\sum_{i\in S(X)}\ind{p_i(X)>\beta k/m} + k-1\right\}\,.
\label{eq:U-Simes}
\end{align}
Here, since the null distribution $P_0$ is assumed to be known, we consider the $p$-values $p_i(X)=\overline{F_0}(X_i),\:\:\:1\leq i\leq m$, where $\overline{F_0}(x)=\P(Z\geq x)$ for any $Z\sim P_0$.
 The bounds considered are summarized in Table~\ref{tab:num-exp-bounds}. Note that for all but the Simes bound, we also have a corresponding lower bound.
\begin{table}[!htp]
  \centering
  \begin{tabular}{lll}
    Name             & Definition            & Reference\\
    \hline
    \textbf{Oracle}  & $U_\beta( S(X),\para)$ & \eqref{equVnew2}\\
    \textbf{Plug-in} & $U_\beta( S(X),\wh{\para)}$ & \eqref{Qplugin}\\
    \textbf{Boot1}   & $\Uboot_{\beta, \delta}( S(\cdot),\wh{\para)}$, $\delta \in \{0.1;0.5;0.9\}$ & Section~\ref{sec:boot1}\\
    \textbf{Boot2}   & $\Ubootbis_{\beta,\delta}( S(X),\wh{\para)}$, $\delta = 0.5$ & Section~\ref{sec:boot2}\\
    \textbf{Naive}   & $\Unnaive_{\beta}(S(\cdot))$ & Section~\ref{sec:boot3}\\
    \textbf{Boot3}   & $\Unrecentred_{\beta}( S(X),\wh{\para)}$ & Section~\ref{sec:boot3}\\
    \textbf{Simes}   & $U^{\textrm{Simes}}(S(X))$ & \eqref{eq:U-Simes}\\
  \end{tabular}
  \caption{(Upper) bounds considered in the numerical experiments}
  \label{tab:num-exp-bounds}
\end{table}

To ensure that the bootstrap bounds are never less conservative than the Plug-in, these bounds have been slightly modified and $\tilde{q}_{\beta}$ has been replaced by its positive part $\tilde{q}_{\beta}^+$. 
 This modification generally has no effect on the upper bounds in practice, but does affect the corresponding lower bounds,
for which we have observed a tendency of the bootstrap towards overcompensation.

Throughout this section, the target risk level is set to $\beta = 0.1$. Therefore, our proposed upper bounds are supposed to  satisfy $\P(U_{\beta}(X,S(X);\para)<\FDP(\theta,S(X))) \leq 0.1$ and the corresponding lower bounds are supposed to satisfy $\P(L_{\beta}(X,S(X);\para)>\FDP(\theta,S(X))) \leq 0.1$.

\paragraph{Selection policies.} We consider three different selection policies $S(\cdot)$, labeled as follows in the figures:
\begin{itemize}
\item \textbf{``$S(X) = \{p_i< 0.05\}$''}: the $p$-value level set associated to the threshold $0.05$. 
\item \textbf{``$S(X) = SC(0.05)$''}: items selected by the procedure introduced by Sun and Cai
  \citep{sun2009large}  in the HMM model (see Section~\ref{sec:relptest}) to control FDR at level $0.05$ using the parameter estimators $\wh{\para}$ of Section~\ref{sec:estim}.
\item \textbf{``$S(X) = \textnormal{Viterbi}$''}: items $i$ selected by the Viterbi algorithm, that is, such that $\widehat{\theta}_i=1$, where $\widehat{\theta}$ is the estimation of $\theta$ using the Viterbi algorithm with the parameter estimators $\wh{\para}$ of Section~\ref{sec:estim}.
\end{itemize}

Section~\ref{sec:results-ident} illustrates the behavior of the considered bounds in a setting where our assumptions  are met. The robustness of the method is then studied in Section~\ref{sec:chall-model-assumpt}, where we report the results of numerical experiments in the case where the HMM is not or poorly identifiable, and where the selected set $S(X)$ depends on prior knowledge not included in $X$. Finally, in Section~\ref{sec:simu-jointseg} we report the results of further numerical experiments on  DNA copy number data from a cancer study with known truth. 

\subsection{Results in a typical within-model scenario}
\label{sec:results-ident}

 The Markov chain $(\theta_i)_{1 \leq i \leq m}$ is generated from transition matrix :
 $$
 A =  \left( \begin{matrix}
 0.95 & 0.05 \\
 0.2  & 0.8 
 \end{matrix} \right).
 $$
with the stationary distribution $(0.8, 0.2)$ as initial state.

 The $X$ variables are generated such that $X_i|\theta_i=0 \sim \mtc{N}(0, 1)$ and $X_i|\theta_i=1 \sim P_1$. Here, $P_1 = \mtc{N}(3, 1)$.
 \begin{figure}
 \begin{tabular}{cc}
 (A)& \includegraphics[width= \textwidth]{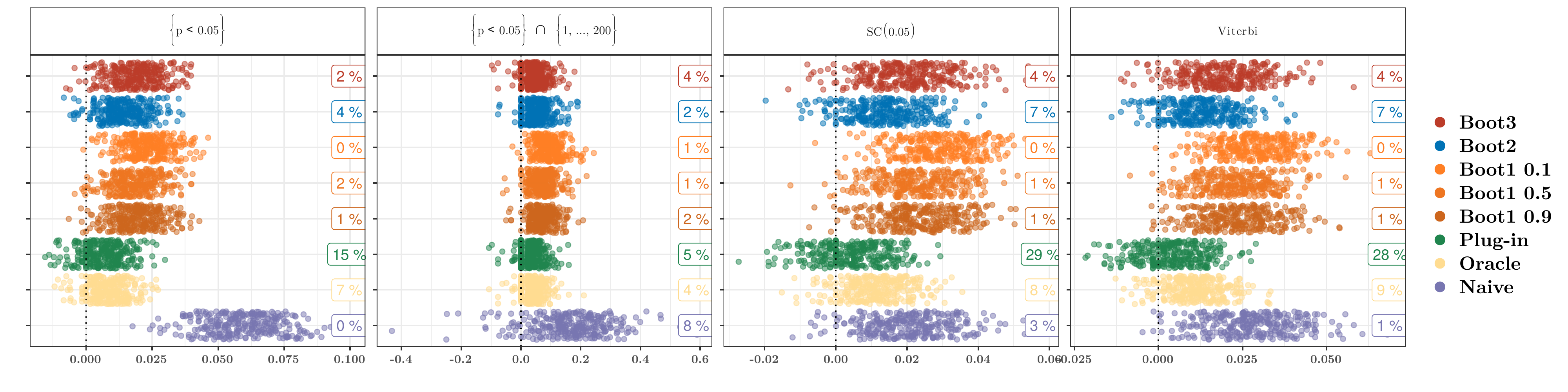}\\
 (B) & \includegraphics[width= \textwidth]{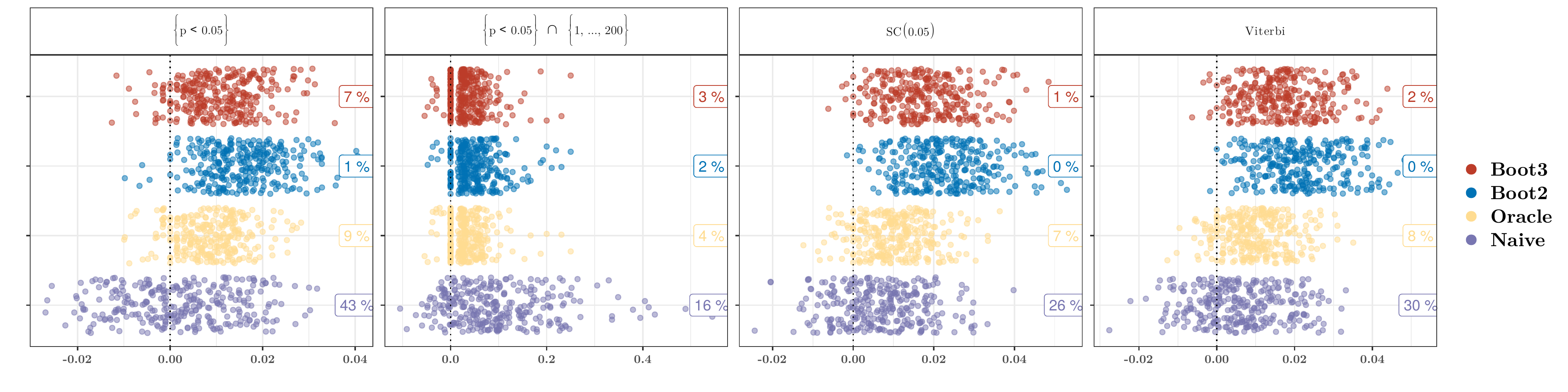}\\
 (C) &  \includegraphics[width= \textwidth]{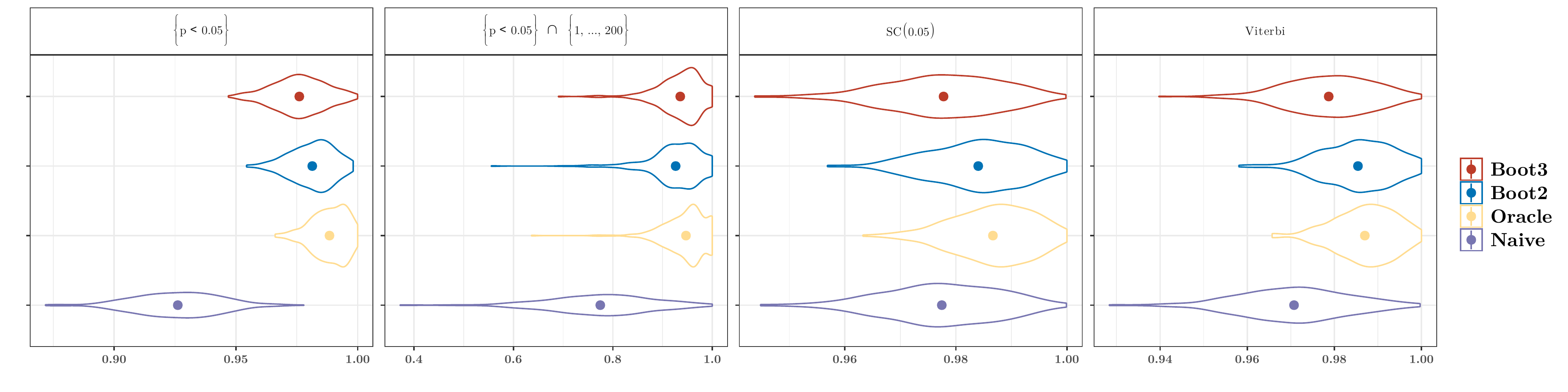} \\
 (D) &  \includegraphics[width= \textwidth]{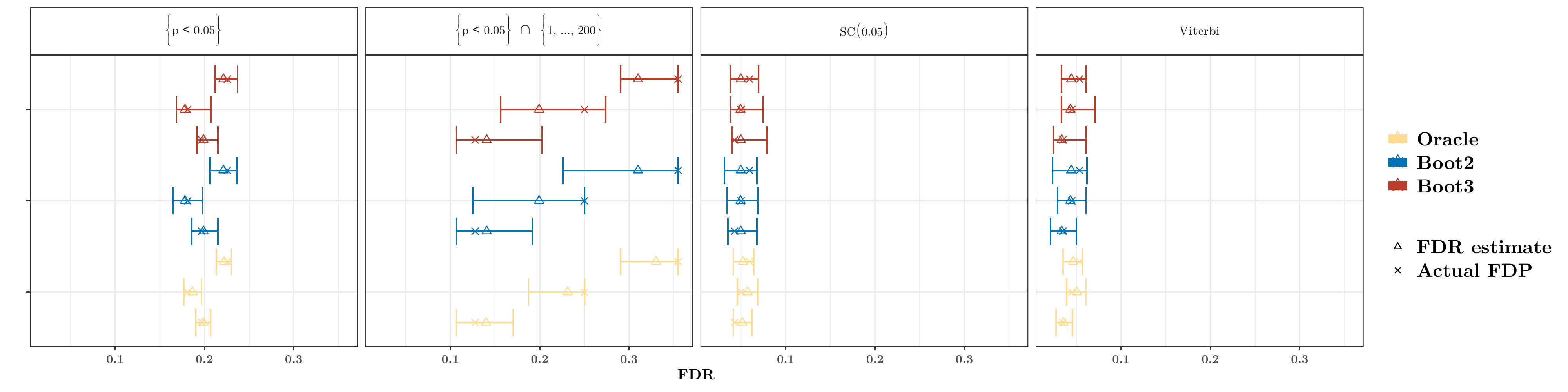}
 \end{tabular}
 \caption{ 
(A-B-C): Results for 300 simulation runs with the model parameters described in Section \ref{sec:results-ident}
(A): Difference $\Delta = U(\cdot)- \FDP(\theta,S(X))$ for each upper bound (rows) and each selection policy $S(X)$ (columns), with the empirical violation probability of the bound displayed within rectangles. 
(B): Difference $\Delta' =\FDP(\theta,S(X))- L(\cdot)  $ for the corresponding lower bounds.
(C): Power of the different bounds.
(D): Realizations of 80\% FDP post hoc intervals for three illustrative simulation runs.
 }
 \label{fig:highdet}
 \end{figure}
 
Figure~\ref{fig:highdet} (A) displays the value of the difference  $\Delta$ between the upper bound $U(\cdot)$ and the true FDP value $\FDP(\theta,S(X))$ for each upper bound (in rows) and selection policy (in columns), for 300 simulation runs. The empirical violation probability of the bound, that is, the proportion of simulation runs for which a given bound is  lower than the true FDP (i.e. $\Delta < 0$), is displayed within rectangles. This proportion is expected to be lower than $\beta = 10 \%$.  We chose not to display the classical Simes bound because, as expected, it is far more conservative than the proposed bounds that take into account the latent structure of the model.  

As stated earlier, the Plug-in bound is an estimator of the oracle bound and not an upper bound of it, hence it is not surprising that its empirical violation probability exceeds the target. 
All of the other bounds have  empirical violation probability below the target in this setting. This is expected for the bootstrap-corrected bounds Boot1, Boot2, and Boot3. The fact that the Naive bound is also below the target risk was not guaranteed, because (similarly to  $U(\beta,S(X);\wh{\para})$) this bound is only an estimator of $U(\beta,S(X);\para)$. 
 We observed that, the impact of the choice of $\delta$ in Boot1 seems moderate. Nevertheless Boot3 seems always less conservative than Boot1, which could be explained by the fact that Boot3 does not use such split of the confidence budget. 
 Overall, Boot2 and Boot3 do not uniformly dominate each other, which is well expected because they come from two different bootstrap strategies.
 For the sake of readability of the illustrations, Boot1 and the Plug-in bound will not be displayed in the remainder of this section.

%
 Figure~\ref{fig:highdet} (B) displays, for the corresponding lower bound the value of the difference $\Delta'$ between the FDP and the lower bound $L(\cdot)$. Similarly to (A), valid lower bounds are expected to be above the true FDP (i.e. $\Delta' <0$) in less than 10\% of the simulations runs.  This is the case for all the proposed bounds, expect for the naive one.
Figure~\ref{fig:highdet} (C) displays the power of the different bounds, which is defined consistently with \cite{blanchard2020post}:
\begin{align}\label{equ:power}
\mbox{Power} = \E\left( \frac{|S(X)| - U(X)}{|S(X)\cap \cH_1|}\:\bigg|\: |S(X)\cap \cH_1|\neq 0, |S(X)\cap \cH_0| \leq  U(X) \right)
\end{align}
In this setting, the bounds are almost as powerful as the oracle.

Finally, in order to emphasize the interest of FDP post hoc interval compared to 
pointwise FDR estimate $\widehat{\FDR}(S(X),\wh{\para})$, see \eqref{equFDRestimate}, we have displayed both in Figure~\ref{fig:highdet} (D) for 3 arbitrarily chosen simulation runs.
The FDP post hoc intervals are clearly more informative than the corresponding point estimate
since they quantify its uncertainty, which can be widely different according to the scenario:
this is reflected in the interval lengths.  In two of the three displayed simulation the estimated FDR is quite far from the true value, whereas the true value still lies in the  post hoc interval.

\subsection{Challenging  our assumptions}
\label{sec:chall-model-assumpt}
This section briefly summarizes further numerical experiments 
carried out in order to test the robustness of the proposed bounds either to
violations of the model assumptions, or to departures from a mild scenario
towards more challenging settings. The corresponding illustrations are postponed
to Appendix~\ref{sec:addit-numer-exper}.

\paragraph{Invalid selection policies.} 
One of the assumptions of the method is that the selected set $S(X)$ cannot
depend on an additional prior knowledge not included in the observation $X$. In
terms of the modeling via the HMM, violating this assumption would correspond to
a selection policy which would have access to ``insider information'' about the
latent configuration vector $\theta$ under one form or another (e.g. an ancillary statistic
  providing some additional information). To assess how important this constraint is
  for the validity of the method,
we have considered three selection
policies that include full knowledge of $\cH_0$ (this of course a very irrealistic situation,
only considered here to illustrate the point).
The results are displayed in
Figure~\ref{fig:knowledge} (Appendix \ref{sec:addit-numer-exper}) and show that
the oracle bound (as well as the second bootstrap bound which tries to mimic the oracle) are
too liberal in this case. The third bootstrap bound, which corrects the plug-in bound to
try to match the FDP, respects the risk. However, it requires knowledge of the full selection
policy $S(.)$. Depending on the situation, this might be realistic (e.g. if
the additional information stems from an ancillary statistic that can also be simulated) or not
(e.g. if the additional information comes from vaguely defined ``insider expert knowledge'').

\paragraph{Identifiability issues.} 
When the latent variables are independent, the model is not identifiable \citep{alexandrovich2016nonparametric,gassiat2016inference}.
When the latent variables are close to  independence (that is, when  $\det{(A)}$ gets close to 0), the model is close to singular.
The bounds obtained for transition matrices $A$ with various values of  
$\det{(A)}$ are displayed in Figure~\ref{fig:diffdet} (Appendix \ref{sec:addit-numer-exper}). As expected, the bounds are too liberal in the independent case. However, they appear to be valid even for small determinants.

\paragraph{Unknown $f_0$.}
Finally, we have also computed the bounds in a case where $f_0$ is unknown. We have considered two options for initializing the estimation of $f_0$ in the corresponding EM-type algorithm, see Algorithm~\ref{algo:EMf0unknown}: (i) using the true $f_0$, and (ii) estimating $f_0$ using local FDR algorithm ~\cite{efron2004large}.
In both cases, the proposed bounds remain below the target risk, as illustrated in Figure~\ref{fig:highdet_uf0} (Appendix~\ref{sec:addit-numer-exper}).

\subsection{Semi-simulated data based on DNA copy numbers}
\label{sec:simu-jointseg}
In order to test the robustness of our methodology we have considered a more realistic scenario, which does not rely on a probabilistic simulation model. Using  the R package \texttt{jointseg} \citep{jointseg1}, we have generated synthetic copy number profiles as proposed (and further described) in \cite{pierre2015performance}: given a number $m$ of loci and a number $K$ of regions, we draw uniformly $K-1$ breakpoint positions, thus defining $K$ regions. Then, we draw $K$ region labels, corresponding to the number of DNA copies for each parent (a.k.a. parental CN). 
Finally, for a region of size $m_k$, we draw $m_k$ samples by resampling from a real DNA copy number data corresponding to this type of region. These data are available in the R package \texttt{acnr} \citep{acnr}, which contain annotated CN profiles from several cancer data sets. Importantly, these data sets correspond to dilution series where tumor and normal cells are mixed in known proportion. Therefore, signal to noise ratio of the corresponding CN profile is implicitly controlled by the fraction of tumor cells. 

We proceed as follows to obtain two groups of samples with known differential regions. First, we generate $n_1 = 50$ CN profiles as described in the preceding paragraph, with the same regions (here, 10 regions). Then we generate $n_2=50$ samples from the same regions, but modify the label of two regions in such a way that this group of samples only has 8 different regions instead of 10. In this setting, $\theta_i = 1$ if the position $i$ belongs to one of the two modified regions and $\theta_i=0$ otherwise.  We then compute Wilcoxon statistics and scale them using their limit law. More precisely, to compare a group 1 of size $n_1$ to a group 2 of size $n_2$ at position $i$ we use  $T^{(s)}_i$
\begin{equation}\label{eq:wilcox}
T^{(s)}_i = \left(T_i - \frac{n_1 n_2}{2}\right) / \sqrt{\frac{n_1n_2(n_1+n_2+1)}{12}},
\end{equation}
where $T_i$ is the classical Wilcoxon statistic, the sum of the rank of group 1. 
 This process is illustrated by Figure~\ref{fig:jointseg-regions} in Appendix~\ref{sec:suppl-plots-jointseg}. In this particular example, the proportion of tumor cells has been set to 70\%, corresponding to a moderate to high signal to noise ratio.  

\begin{figure}
\includegraphics[width=\textwidth]{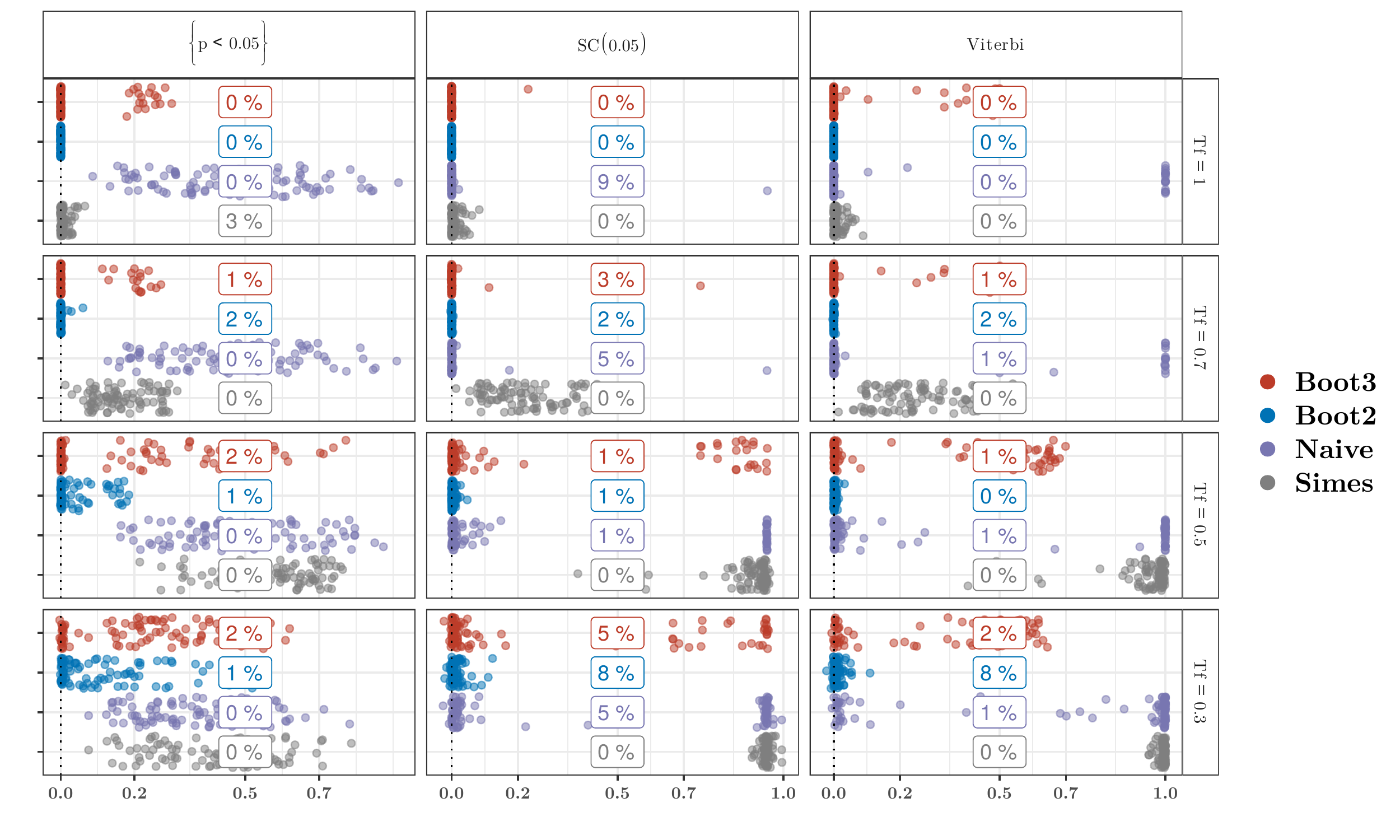}
\caption{
Summary of 100 numerical experiments with semi-simulated CN data. Each dot corresponds to a realization of the difference $\Delta = U(\cdot)- \FDP(\theta,S(X))$ for each upper bound (colors) and each selection policy $S(X)$ (columns) for different values of the SNR (rows). The  empirical violation probability of the bound is displayed within a rectangle.}
\label{fig:real}
\end{figure}

The results of these numerical experiments are summarized in Figure~\ref{fig:real}, where the difference  $\Delta$ between the FDP upper bound  and the true FDP is displayed for each of 100 simulation runs. The lines in this figure  correspond to increasing values for the  SNR (as governed by the fraction of tumor cells), while the columns correspond to the selection policies introduced at the beginning of Section~\ref{sec:num}. 

All the bounds appear to be valid in this settings, since their empirical violation probability is less than the target level $\beta = 10\%$ (i.e. $\Delta < 0$ in less than 10\% of the simulation runs). Overall, the bounds are quite conservative in all settings, with empirical  violation probabilities often closer to 0 than $10\%$. This conservativeness is partly explained by the fact that for high SNR values, the problem is so easy that the upper bounds often match the true FDP exactly, as illustrated by the presence of a mode at 0. The Simes bound is the most conservative, with null empirical  violation probability in all but one scenario. 
As SNR decreases, we observe for all bounds a shift of the empirical distribution of $\Delta$ toward positive values.  The Simes bounds is highly conservative already at tumor fraction 0.7, and $\Delta$ even concentrates at $1$ for three of the four selection policies at tumor fractions 0.3 and 0.5: this corresponds to a true FDP close to 0 and an upper bound close to 1, meaning that the Simes bound is not informative at all in these settings. The first boostrap bound has the same tendency, but to a lesser extent. The second and third bootstrap bounds show a remarkable behavior, with $\Delta$ remaining very close to 0 for the vast majority of simulation runs for all selection policies, even for small SNR values. 
In addition, Boot2 and Boot3 have also a high power \eqref{equ:power}, as shown in Figure \ref{fig:reali-power}
 (Appendix~\ref{sec:powerrealist}).
These results strengthen our confidence in the applicability of our methods and that even in practical applications, our proposed bounds (and especially the second and third bootstrap bounds) will be able to evaluate the amount of true signal after selection with accuracy and correct coverage.


%% file: appli.tex
\subsection{Influenza-like illness (ILI)}
\label{sec:infl-like-ill}
We apply the proposed method to the weekly incidence rates of influenza-like illness (ILI). This data were collected from the Sentinelles Network, a national surveillance system in France.
\cite{sun2009large} studied this data set between January 1985 and February 2008, to be comparable we will restrict ourselves to this period. 
They  stated that the incidence rates can be classified into one of the two categories: aberration or usual. \\
The usual is the null states and the aberration is the alternatives one. 
The weekly ILI incidence rates are standardized according to the sizes of the underlying population and the representativeness of the participating physicians. 
\cite{sun2009large}  also applied a log transformation to reduce the skewness of the original data, hence so will we.
In this particular example, the law under the null hypothesis ($P_0$) is unknown. Therefore, we estimate $\wh{f}_0$ using Algorithm~\ref{algo:EMf0unknown}, with $\mathcal{N}(2.37, 0.76^2)$, the estimation of \cite{sun2009large}, as initial value. Accordingly, the $p$-values are replaced by the empirical $p$-values $\wh{p}$ defined in Appendix~\ref{sec:suppl-unknown-f0}.

 To emphasize the advantages of bounds on the  FDP compared to the FDR {estimate}, we displayed the $80 \%$ FDP post hoc intervals on the different sets $S(X)$ described in  the numerical experiments. 
For instance, for the selection $SC(0.05)$, it is interesting to know not only that the estimated FDR is $5\%$ but also that the $90\%$ upper bound is smaller than $0.08$. 
{ This application also underlines the interest of developing lower bounds, for instance in the first set ($\{\wh{p} < 0.05\}$) we not only know, with high probability, that the FDP is smaller than 0.054 but also that it is higher than 0.028, which narrows down the probable true value of the FDP.}

\begin{figure}
\center
\includegraphics[width = 0.6\textwidth]{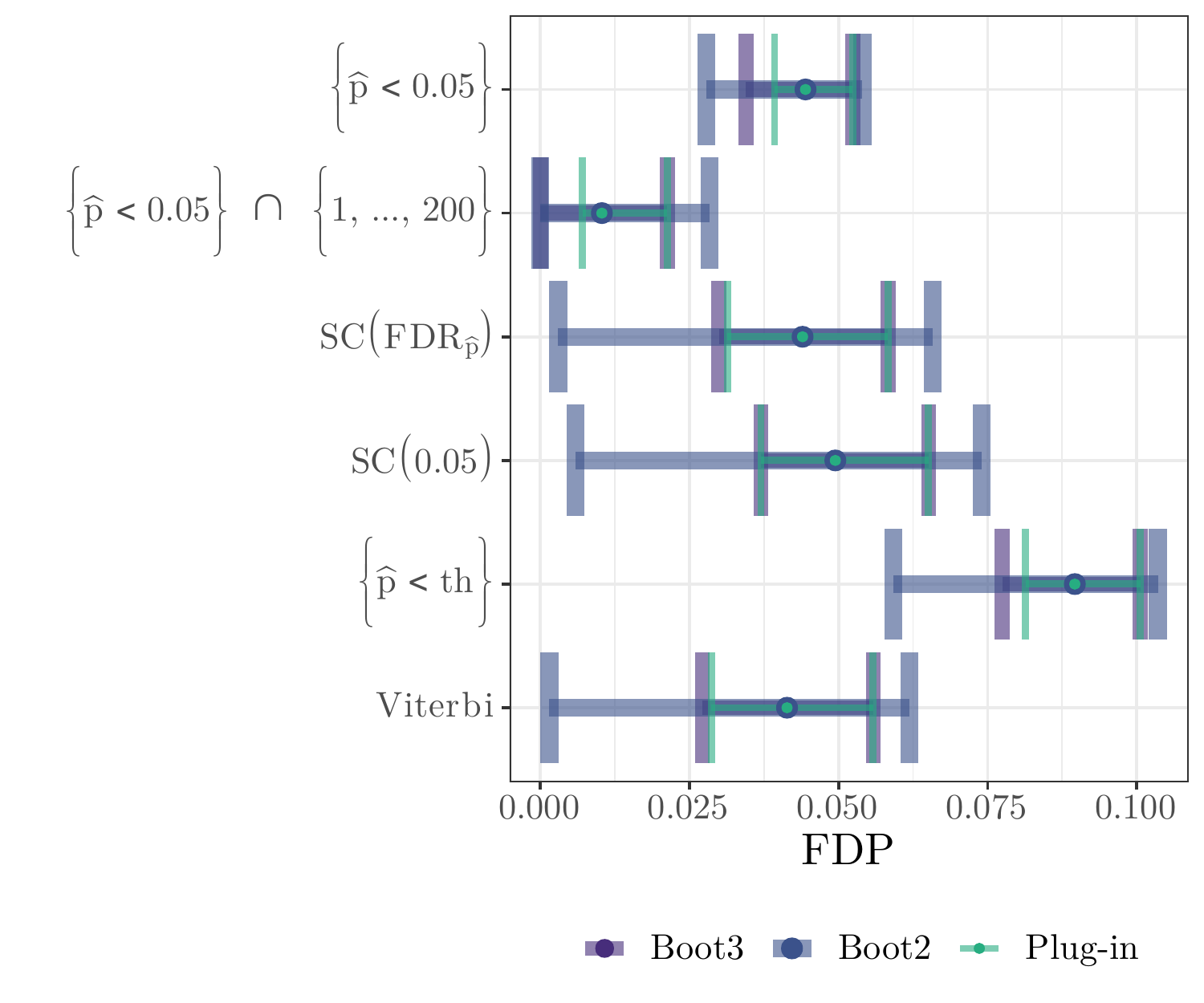}
\caption{80 \% FDP post hoc interval  intervals for different selection policies on the ILI incidence rates}
\label{fig:ili2}
\end{figure}

\marie{
For completeness, we also added two selections policies :
 \begin{itemize}
 \item $SC(FDR_p)$ which uses \cite{sun2009large} algorithm (with our parameter estimator) at a level corresponding to the estimated FDR of the set $\{\wh{p}<0.05\}$. Doing so, the selections $SC(FDR_p)$ and $\{\wh{p}<0.05\}$ have the same estimated FDR.
  \item $\{\wh{p} < th\}$ which selects the $|SC(0.05)|$ smallest $\wh{p}$-values. Doing so, $\{\wh{p} < th\}$ and $SC(0.05)$ select the same number of null hypotheses.
 \end{itemize}
 We added the selection policies $SC(FDR_p)$  to compare two selection sets that have the same FDR estimate but not the same size. 
 The selection $\{\wh{p} < th\}$ has been added to compare two sets that have the same size but not the same estimated FDR. 
 The result are displayed Figure~\ref{fig:ili2}.
 All the intervals are sharp around the FDP, which means that the FDP variance is estimated to be low. 
However, the post hoc intervals of $SC(FDR_p)$ is wider than the one of  $\{\wh{p}<0.05\}$. This information is not provided by FDR estimates.
The set $\{\wh{p} < th\}$ has the same size that  $|SC(0.05)|$ but a larger FDR. It emphasizes that \cite{sun2009large} is a better selection policy.
 Overall, this reinforces the interest in using a post hoc interval rather than a pointwise FDR estimate.
}

 \subsection{Copy Number Alterations Associated with  Endometriosis in Ovarian Clear Cell Adenocarcinoma}
\label{sec:copy-numb-alter}

\begin{figure}
\begin{tabular}{ccc}
(A) &\includegraphics[width=0.5\textwidth]{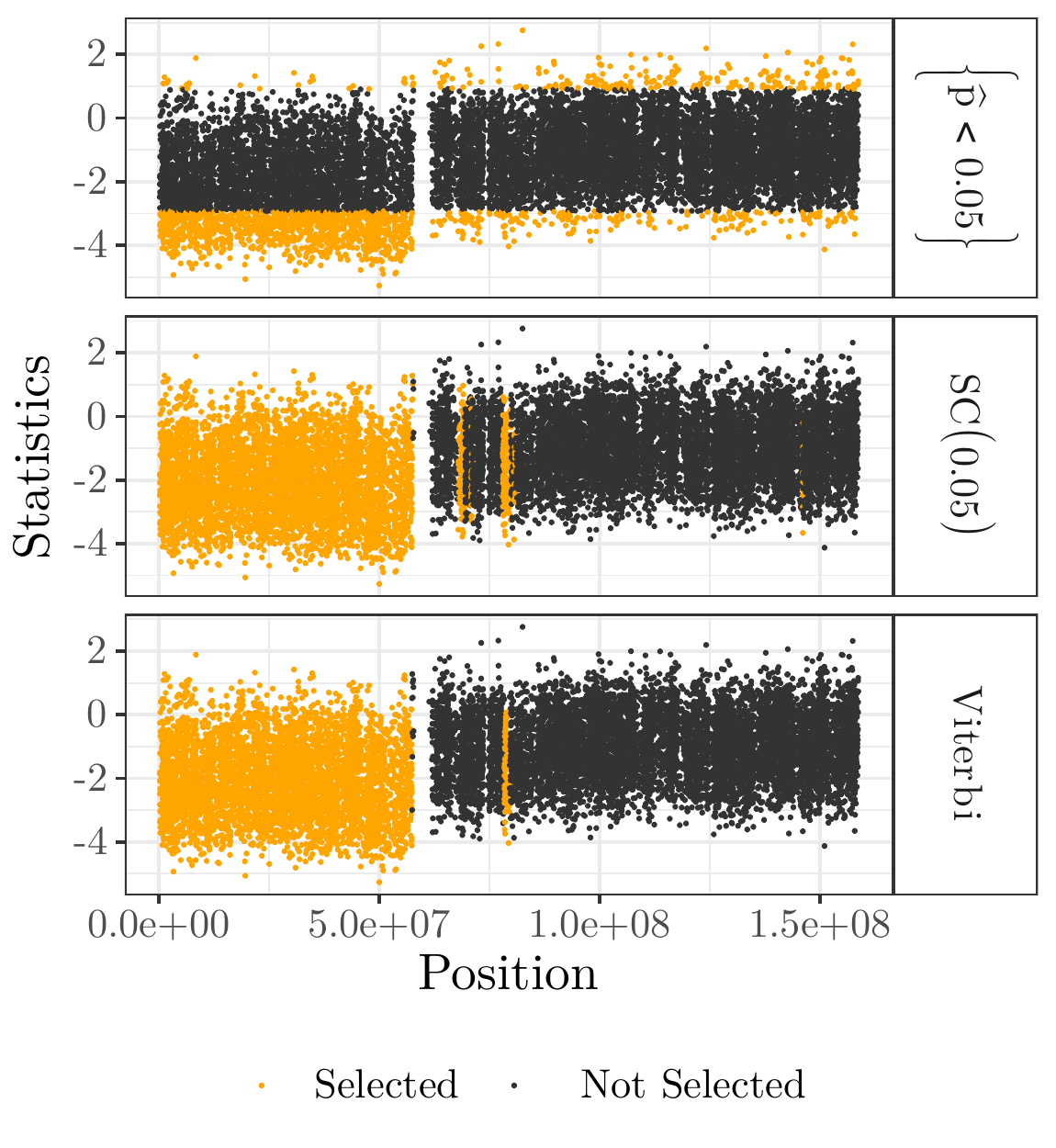} & 
\includegraphics[width=0.5\textwidth]{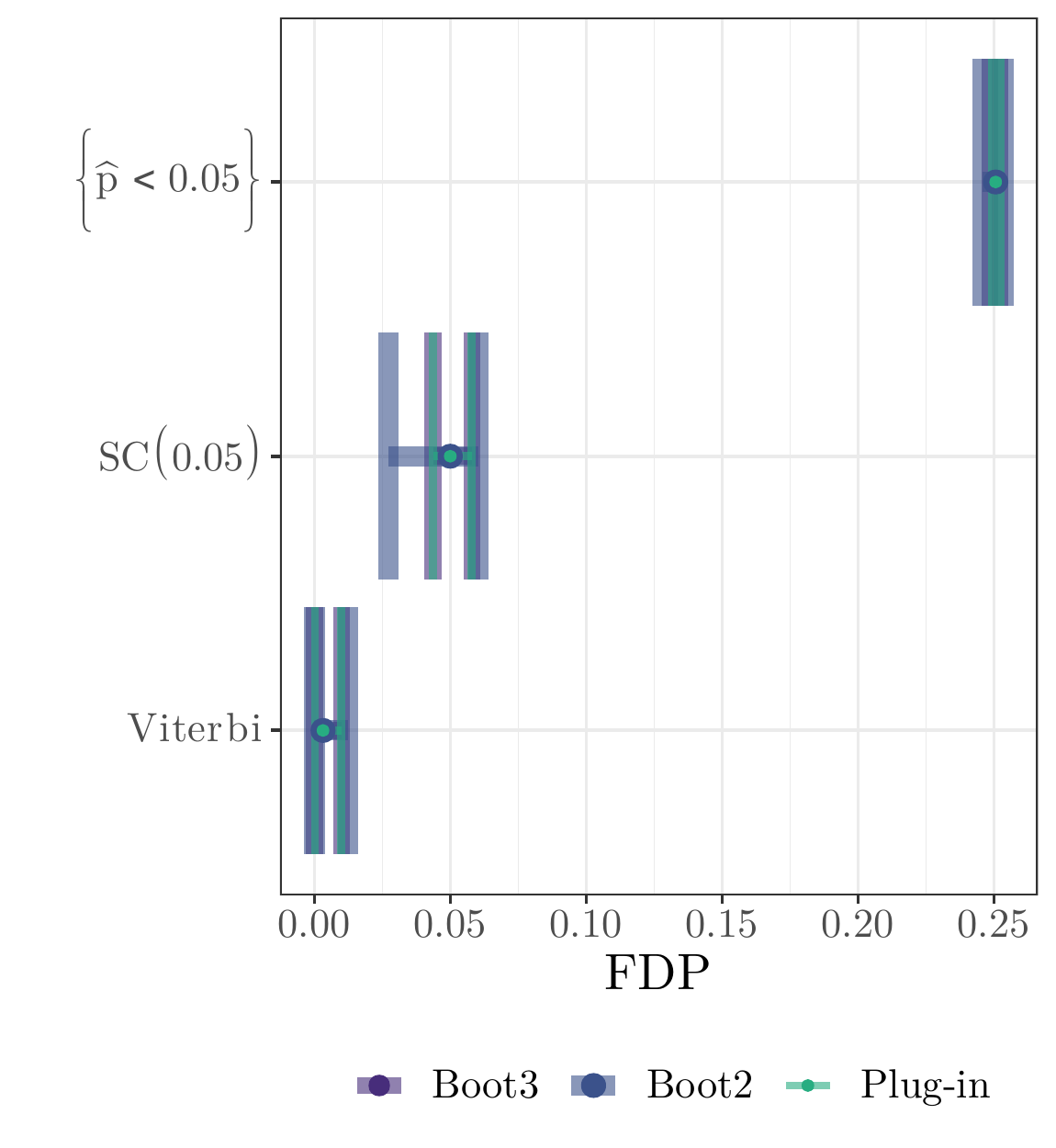} \\
(B) & \includegraphics[width=0.5\textwidth]{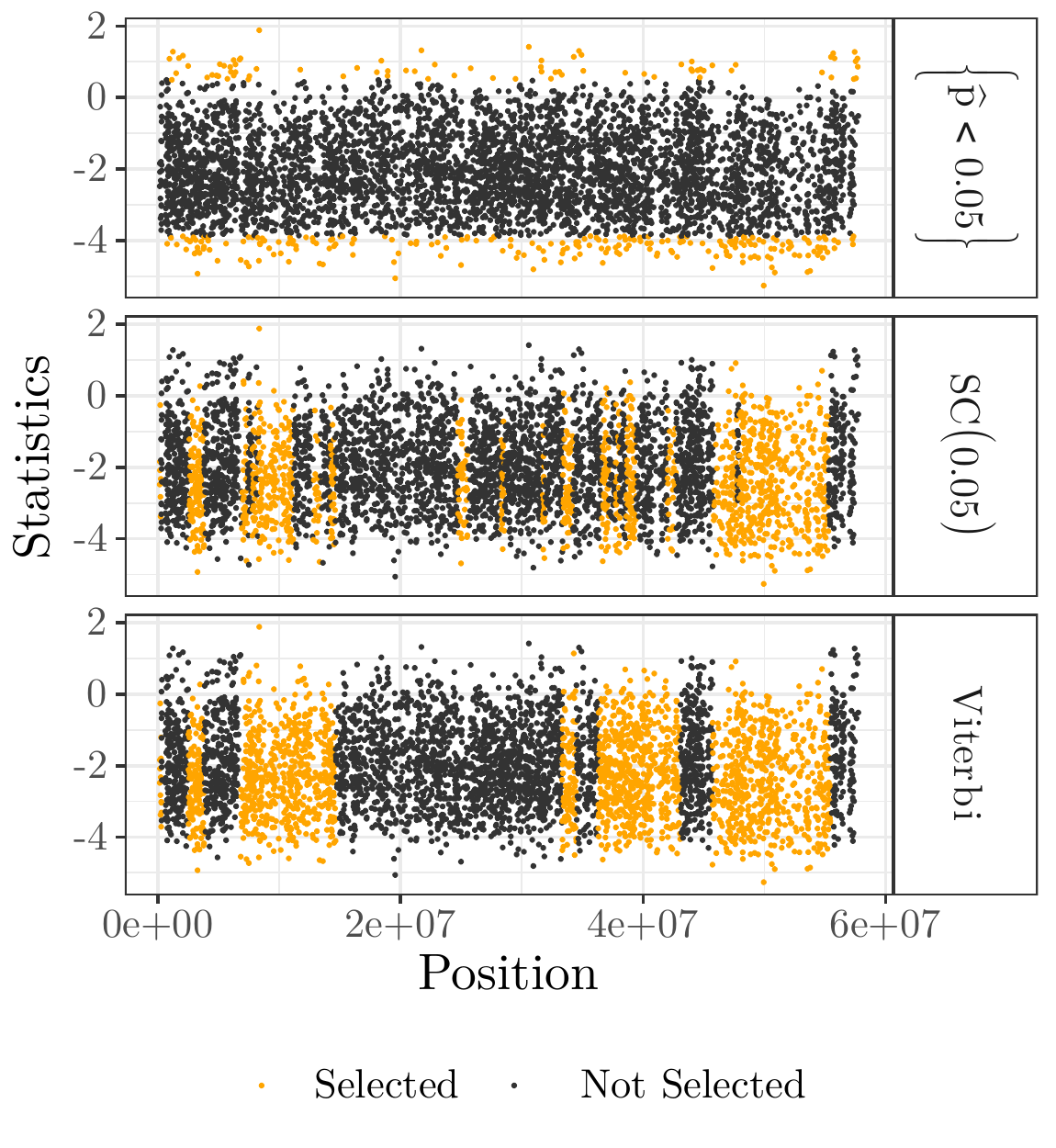} & 
\includegraphics[width=0.5\textwidth]{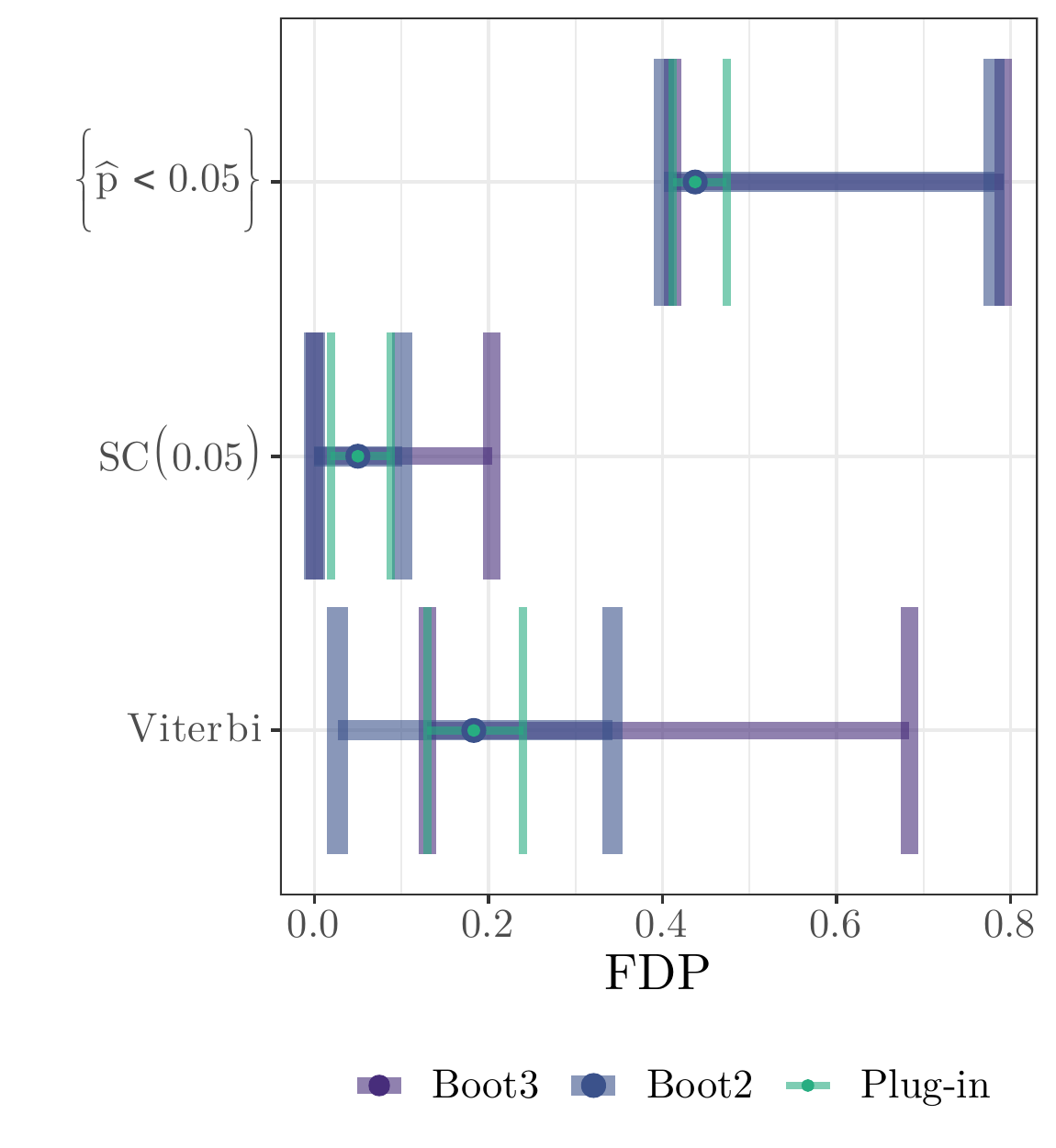}
\end{tabular}
\caption{Left : Wilcoxon test statistic $T_i^{(s)}$ (see Equation~\ref{eq:wilcox}) comparing the copy numbers between patients with and without endometriosisat different loci on chromosome 7 for (A) and 7p for (B). Right : 80\% FDP post hoc interval different selection policies. Since $f_0$ is unknown, empirical $p$-values defined in  Appendix~\ref{sec:suppl-unknown-f0} are used.}
\label{fig:full7} 
\end{figure}

In this section our proposed bounds are applied to a study of 117 ovarian cancer
patients briefly mentioned in the introduction~\citep{okamoto2015somatic}.  We
focus on the part of the study that aims at comparing, for 13,239 loci located on 
chromosome 7, the DNA copy numbers of 54 patients with endometriosis (a common gynecological disorder characterized by ectopic growth of endometrial glands and stroma) to that of 63 patients without endometriosis.  
As in the semi-simulated setting described in Section~\ref{sec:simu-jointseg} we compare the two groups using Wilcoxon tests and scale the statistics using Equation~\eqref{eq:wilcox}. The obtained statistics for the different positions of the genome are displayed in the left part of Figure~\ref{fig:full7} (A) for the full chromosome 7 and (B) for the ``short arm'' of chromosome 7 (corresponding to the first 4,799 loci). 

\pierre{
A first important observation is that the test statistics are not centered at 0 but globally shifted toward negative values across the entire chromosome 7, and more prominently so in chromosome 7p. This indicates that patients with endometriosis have on average a larger CN than the others, which may be due to an increase prevalence of trisomy 7 in these patients, or to a larger proportion of tumor cells in the biological samples corresponding to these patients. Here, our goal is not to detect such macroscopic changes but to pinpoint chromosomal regions that deviate from the rest of the chromosome. Indeed, such regions could indicate the presence of ``driver'' genes more actively or more early implicated in the tumor process compared to other ``passenger'' regions.}

\pierre{
Therefore, we chose to estimate $f_0$ from the data, as described in Appendix~\ref{estimf0}. Note that $\wh{f}_0$ will not estimate the distribution of the null hypothesis ``the two groups have the same number of DNA copy at this position'' but the distribution of the most frequent type of difference between the two groups. For instance if all patients without endometriosis have two copies of chromosome 7 (not affected by their ovarian cancer) and patients with endometriosis have a third copy of chromosome 7 and in some rare position they have a fourth copy, the estimated $f_0$ will be the law of the statistics comparing copy numbers two and three. 
Because of the marked shift between the short (left) and long (right) arms of chromosome 7, the estimation of $f_0$ is quite different depending on whether the entire chromosome or only one arm is analyzed. Therefore, we have carried out an analysis of the entire chromosome (Figure~\ref{fig:full7} (A)) and an analysis of its short arm, chromosome 7p  (Figure~\ref{fig:full7} (B)):}

\pierre{
\begin{description}
\item[Full chromosome 7:] By construction, the selection policies that take into account the position mostly select regions on chromosome 7p.
The post hoc intervals corresponding to each selection policy are represented in the right panel of Figure~\ref{fig:full7} (A). In this scenario the post hoc intervals are very tight around the pointwise FDR estimate, which reflects the fact that $\hat{f}_0$ and $\hat{f}_1$ are quite distinct here. 
\item[Chromosome 7p:]
When focusing on chromosome 7p,  $\hat{f}_0$ and $\hat{f}_1$ are much closer to each other, resulting in wider FDP post hoc intervals than for the full chromosome. For instance, in the sets SC(0.05) the FDR is controlled at 5\% but the upper bounds goes up to 20\%. Even the plug-in bound, which has been shown in the simulation not to be conservative enough, goes up to 10\%.
\end{description}
}
  
\pierre{
These examples emphasize the added value of post hoc intervals on the FDP over FDR point estimates, since a given value of the point estimate will be interpreted very differently depending on the width of the interval.
}
